\documentclass[twocolumn,iop]{emulateapj}

\usepackage{graphics,color}
\graphicspath{{Figures/}{./}}
\usepackage{amsmath}

\usepackage{txfonts}
\usepackage[breaklinks=false]{hyperref} 
\usepackage{color}
\usepackage{sistyle}

\SIthousandsep{,}

\newcommand{\hMpc}{h^{-1}~\mathrm{Mpc}}

\begin{document}

\title{
The extended Baryon Oscillation Spectroscopic Survey:
measuring the cross-correlation between the MgII flux transmission field and
quasars and galaxies at $z=0.59$
}

\slugcomment{Submitted to ApJ}

\author{
H\'elion~du~Mas~des~Bourboux\altaffilmark{1},
Kyle~S.~Dawson\altaffilmark{1},
Nicol\'as~G.~Busca\altaffilmark{2}, 
Michael~Blomqvist\altaffilmark{3},
Victoria~de~Sainte~Agathe\altaffilmark{2},
Christophe~Balland\altaffilmark{2},
Julian~E.~Bautista\altaffilmark{4},
Julien~Guy\altaffilmark{5},
Vikrant~Kamble\altaffilmark{1},
Adam~D.~Myers\altaffilmark{6},
Ignasi~P{\'e}rez-R{\`a}fols\altaffilmark{3},
Matthew~M.~Pieri\altaffilmark{3},
James~Rich\altaffilmark{7},
Donald~P.~Schneider\altaffilmark{8,}\altaffilmark{9},
An\v{z}e~Slosar\altaffilmark{10}
}

\altaffiltext{1}{
Department of Physics and Astronomy, University of Utah, 115 S 1400 E, Salt Lake City, UT 84112, USA
}
\altaffiltext{2}{
Sorbonne Universit\'e, Universit\'e Paris Diderot, CNRS/IN2P3,\\Laboratoire de Physique Nucl\'eaire et de Hautes Energies, LPNHE, 4 Place Jussieu, F-75252 Paris, France
}
\altaffiltext{3}{
Aix Marseille Univ, CNRS, LAM, Laboratoire d'Astrophysique de Marseille, Marseille, France
}
\altaffiltext{4}{
Institute of Cosmology \& Gravitation, University of Portsmouth, Dennis Sciama Building, Portsmouth, PO1 3FX, UK
}
\altaffiltext{5}{
Lawrence Berkeley National Laboratory, 1 Cyclotron Road, Berkeley, CA 94720, U.S.A
}
\altaffiltext{6}{
Department of Physics and Astronomy, University of Wyoming, Laramie, WY 82071, USA
}
\altaffiltext{7}{
IRFU, CEA, Universit\'e Paris-Saclay,  F-91191 Gif-sur-Yvette, France
}
\altaffiltext{8}{
Department of Astronomy and Astrophysics, The Pennsylvania State University, University Park, PA 16802
}
\altaffiltext{9}{
Institute for Gravitation and the Cosmos, The Pennsylvania State University, University Park, PA 16802
}
\altaffiltext{10}{
Brookhaven National Laboratory, Physics Department, Upton, NY 11973, USA
}

\email{h.du.mas.des.bourboux@utah.edu}

\shorttitle{BAO with MgII}

\begin{abstract}
We present the first attempt at measuring the baryonic acoustic oscillations (BAO) in the
large scale cross-correlation between the magnesium-II doublet (MgII) flux transmission field and
the position of quasar and galaxy tracers.
The MgII flux transmission continuous field at \mbox{$0.3 < z < 1.6$} is measured
from \num{500589} quasar spectra obtained in the
Baryonic Oscillation Spectroscopic Survey (BOSS) and the extended BOSS (eBOSS).
The position of \num{246697} quasar tracers and \num{1346776}
galaxy tracers are extracted from the Sloan Digital Sky Survey (SDSS) I, II, BOSS and eBOSS catalogs.
In addition to measuring the cosmological BAO scale and the biased matter density
correlation, this study allows tests and improvements to cosmological Lyman-$\alpha$
analyses.
A feature consistent with that of the BAO is detected at a significance of
$\Delta \chi^{2} = 7.25$.
The measured
MgII linear transmission bias parameters are
$b_{\mathrm{MgII(2796)}} (z = 0.59) = (-6.82 \pm 0.54) \, \times 10^{-4}$
and
$b_{\mathrm{MgII(2804)}} (z = 0.59) = (-5.55 \pm 0.46) \, \times 10^{-4}$,
and the MgI bias is
$b_{\mathrm{MgI(2853)}}  (z = 0.59) = (-1.48 \pm 0.24) \, \times 10^{-4}$.
Their redshift evolution is characterized by the power-law index:
$\gamma_{\mathrm{Mg}} = 3.36 \pm 0.46$.
These measurements open a new window towards using
BAO from flux transmission at $z < 1$ in the final eBOSS sample and in the upcoming sample
from the Dark Energy Spectroscopic Instrument.
\end{abstract}

\hypersetup{pdftitle={
The extended Baryon Oscillation Spectroscopic Survey:
measuring the cross-correlation between the MgII flux transmission field and
quasars and galaxies at \texorpdfstring{$z=0.59$}{z=0.59}
}}
\hypersetup{pdfsubject=Cosmology}
\hypersetup{pdfauthor={H. du Mas des Bourboux et al.}}
\hypersetup{pdfkeywords={cosmology, distance scale, large-scale structure of universe, quasar, intergalactic medium, absorption lines}}

%
%
\section{Introduction}
\label{section::Introduction}

The intergalactic medium (IGM) gas traces the underlying distribution
of baryonic matter and dark matter. In spectra of background quasars,
the fluctuations in density of the IGM are observed as a continuous
field of absorption with respect to the unabsorbed emission of the object
\citep{1965ApJ...142.1633G, 1971ApJ...164L..73L}. Different atomic transitions are used to trace
these density fluctuations. The Lyman-$\alpha$ (Ly$\alpha$) transition
from the first orbital to the second orbital of the hydrogen atom produces
the strongest signal. The continuum of absorption from Ly$\alpha$, tracing
the overall fluctuations of matter density is called the Ly$\alpha$ forest.
For redshift $z > 2$, it can be observed from ground-based instruments.
To probe lower redshifts with this technique,
because of atmospheric UV cut-off,
it is necessary either to observe quasars from
space-based instruments \citep[e.g.,][]{1993ApJS...87....1B, 2018arXiv180805605K}
or to use weaker metal
transitions, as suggested by \citet{2014MNRAS.445L.104P}, such as
singly-ionized magnesium \citep[magnesium-II; e.g.,][]{2015MNRAS.447.2784P}
or triply-ionized carbon \citep[carbon-IV; e.g.,][]{2018JCAP...05..029B, 2018MNRAS.480..610G}.

Tracers of the total matter density field are used in cosmology to measure the
biased 3D correlation of matter, host of the baryon acoustic oscillations
(BAO), first detected in galaxies \citep{2005ApJ...633..560E,2005MNRAS.362..505C}.
This latter feature is used as a probe of the
cosmic expansion history. At lower redshift ($z<2$) the BAO scale has been
measured using galaxies
\citep{
2007MNRAS.381.1053P,
2010MNRAS.401.2148P,
2011MNRAS.415.2892B,
2011MNRAS.416.3017B,
2012MNRAS.426..226C,
2012MNRAS.427.2132P,
2012MNRAS.427.2168M,
2013MNRAS.431.2834X,
2012MNRAS.427.3435A,
2014MNRAS.439...83A,
2014MNRAS.441...24A,
2015MNRAS.449..835R,
2017MNRAS.470.2617A,
2018ApJ...863..110B}
and quasars \citep{2018MNRAS.473.4773A}.
At larger redshift ($z>2$) the number density of visible objects declines
drastically and thus the measurement has been made through the Ly$\alpha$
forests auto-correlation
\citep{
2013A&A...552A..96B,
2013JCAP...04..026S,
2013JCAP...03..024K,
2015A&A...574A..59D,
2017A&A...603A..12B}
and through the
Ly$\alpha$-quasar cross-correlation
\citep{
2014JCAP...05..027F,
2017A&A...608A.130D}.

The two methods of extracting the BAO scale differ in technique and
possible sources of systematic errors.
One method uses the position of galaxies or quasars as discrete tracers of
the denser regions of the matter density field, the other uses the
IGM absorption as a continuous tracer of the entire matter density field
along the line-of-sight of a quasar. BAO measurements via these two
methods at the same redshift would enable different systematic tests. As yet
this comparison has not been accomplished, although some steps in this direction have been
investigated. \citet{2016JCAP...11..060L} measured the 3D auto-correlation
of quasars at $z>2$, but do not report a measurement of BAO.
\citet{2018JCAP...05..029B} measured the 3D cross-correlation between the 
carbon-IV (CIV)
absorption in quasar spectra and the quasar distribution with a large
fraction of data at $z<2$; however, they lack a detection of the BAO scale
of comparable precision to discrete tracers.
\mbox{\citet{2015MNRAS.447.2784P}} measured the small scale cross-correlation
between \mbox{magnesium-II} absorbers (MgII) and galaxies at $z \approx 0.5$, but
did not investigate separations
larger than $10 \, \mathrm{Mpc}$.

This study uses the MgII absorption observed in
background quasar spectra as a continuous tracer of the matter density field
to measure the 3D cross-correlation
with galaxies and quasars. Singly ionized magnesium, MgII,
traces metal-enriched, photo-ionized
gas in the circumgalactic medium of galaxies.
In the range of optical spectroscopy with sufficient UV strength
and isolation from strong atmospheric emission,
$3600 < \lambda < 7235 \, \text{\AA{}}$,
MgII covers a large redshift range: \mbox{$0.29 < z < 1.59$}.
In this interval, multiple BAO measurements have been reported from
galaxies \citep[e.g.,][]{2017MNRAS.470.2617A}, allowing for possible comparisons
between discrete and continuous samples of the matter density field.
Our analysis treats the MgII absorption as a continuous field as in
\citet{2014MNRAS.439.3139Z} and in \citet{2015MNRAS.447.2784P},
instead of as a catalog of discrete tracers as done by
\citet{2009ApJ...702...50G} and \citet{2009ApJ...698..819L}.
Treating MgII as a discrete tracer yields a MgII bias of order unity.
\citet{2018JCAP...05..029B} studied CIV as an absorption continuous field;
they measured $b_{\mathrm{CIV}}(z=2) \approx -1.4 \times 10^{-2}$ for the
effective bias of the CIV doublet transition.

The benefits of treating MgII as continuous absorption are:
1) there is no need to identify individual absorbers,
2) there is no confusion with other doublets (e.g., CIV, SiIV) when cross-correlating
with quasars or galaxies, and
3) there is no need to build a catalog of randoms and masks of the
selected spectroscopic targets.
However, the main drawback of this approach is to mix signal from a small number of pixels
(spectral data point of a given wavelength width)
with strong MgII absorption, with numerous pixels without significant absorption.
This technique has the consequence
of producing a low bias compared to discrete MgII bias.

This approach of treating MgII as a continuous tracer is analogous to how Ly$\alpha$
is treated in BAO studies \citep[e.g.,][]{2017A&A...603A..12B}, thus allowing us to test the
Ly$\alpha$ analyses methodology in different regimes.
\begin{itemize}

    \item The Ly$\alpha$ transition is a singlet, while the MgII transition
    is a doublet composed of
    MgII(2796): $\lambda_{\mathrm{R.F.}} = 2796.35 \, \text{\AA{}}$
    and of
    MgII(2804): $\lambda_{\mathrm{R.F.}} = 2803.53 \, \text{\AA{}}$.
    The measured 3D cross-correlation with galaxies or with quasars
    is therefore the superposition of two correlations
    separated by $7 \, \text{\AA{}}$ ($\sim 9 \, \hMpc{}$ at $z=0.59$) and of slightly different
    bias.
    This scenario allows a test of the Ly$\alpha$ analyses in the regime where multiple
    extra correlations are superimposed.

    \item The MgII bias is orders of magnitude lower than that of
    Ly$\alpha$. Systematic errors linked to, for example,
    residuals of the sky subtraction
    or flux calibration, would be more important in
    a correlation involving MgII than Ly$\alpha$.

    \item The MgII absorption is visible down to $z=0.29$, thus enabling
    a cross-correlation of the absorption field with both quasars and galaxies.
    The Ly$\alpha$ field is not visible at $z<2$ in optical spectra, thus it can only be
    cross-correlated with quasars in current spectroscopic samples.
    MgII allows a comparison of the two
    tracers (galaxy and quasars) and a test of possible systematic errors
    associated with the different discrete tracers.

    \item The shape and variation of shape of the different MgII forests
    (sec.~\ref{subsection::measurement_of_the_MgII_flux_fransmission_field})
    differs from the shape of the Ly$\alpha$ forest. This trait allows 
    a search for a source of systematic errors arising from the quasar continuum.
    
    \item As discussed in \citet{2017A&A...603A..12B}, one possible source
    of systematic errors in the Ly$\alpha$ forest auto-correlation is the
    unavoidable presence of all auto-correlations of the different
    metal-absorption features. An independent measure of the bias of the MgII doublet would allow
    a better estimate of this systematic error.

\end{itemize}

We report measurements of the baryonic acoustic oscillations in the 3D cross-correlation
of MgII and galaxies or quasars. In
section~\ref{section::Data_samples_and_reduction}, we present the
catalogs of galaxies and quasars as tracers of matter density fluctuations
and the catalog of quasars as background to the MgII absorption.
We also detail the analysis to measure the absorption fluctuations against
estimates of the unabsorbed quasar continuum.
In section~\ref{section::The_one_dimensional_pixel_auto_correlation},
we study the different metal transitions that contaminate our measurement
using the auto-correlation of pixels from the same background quasar.
Section~\ref{section::The_MgII_quasar_and_MgII_galaxy_cross_correlation}
presents how we measure the cross-correlation between MgII absorptions and quasars
or galaxies and we report the measured correlation functions.
In section~\ref{section::Fit_for_cosmological_correlations}, we describe
the fit to the measured cross-correlations and the resulting measurement
of MgII bias and BAO parameters.
In section~\ref{section::Summary_and_conclusions}, we finish with a
summary and conclusion.

%
%
\section{Data samples and reduction}
\label{section::Data_samples_and_reduction}

The study presented here uses data from the Sloan Digital Sky Survey
\citep[SDSS:][]{2000AJ....120.1579Y}. Most of the tracer quasars, tracer galaxies,
and the entirety of the background quasars were gathered during SDSS-III by the
Baryon Oscillation Spectroscopic Survey
\citep[BOSS:][]{2011AJ....142...72E,
2013AJ....145...10D}, and during
SDSS-IV by the extended BOSS \citep[eBOSS:][]{2016AJ....151...44D, 2017AJ....154...28B}. A small
fraction of tracers were observed during SDSS-I and II.
These data are publicly available in the fourteenth data release
\citep[DR14:][]{2018ApJS..235...42A},
and in the seventh data release \citep[DR7:][]{2009ApJS..182..543A}. All of
these data were acquired with the $2.5$~m Sloan Foundation telescope
\citep{2006AJ....131.2332G} at the Apache Point Observatory.

The catalog of quasar tracers is taken from the DR14 quasar catalog (DR14Q), presented in
\citet{2018A&A...613A..51P}.
The catalog of galaxy tracers is a combination of three different
catalogs:
Luminous Red Galaxies (LRG) from eBOSS \citep{2018ApJ...863..110B},
LRGs from BOSS \citep{2016MNRAS.455.1553R} and
galaxies from SDSS DR7, mainly
from the main sample \citep{2005AJ....129.2562B}.
All quasar spectra used to measure the MgII absorption field
were obtained using the BOSS spectrographs
\citep{2013AJ....146...32S}, which have a spectral
resolution of $\approx 2000$.

\subsection{Catalog of quasars and galaxies}
\label{subsection::Catalog_of_quasars_and_galaxies}

\begin{figure*}
    \centering
    \includegraphics[width=.90\textwidth]{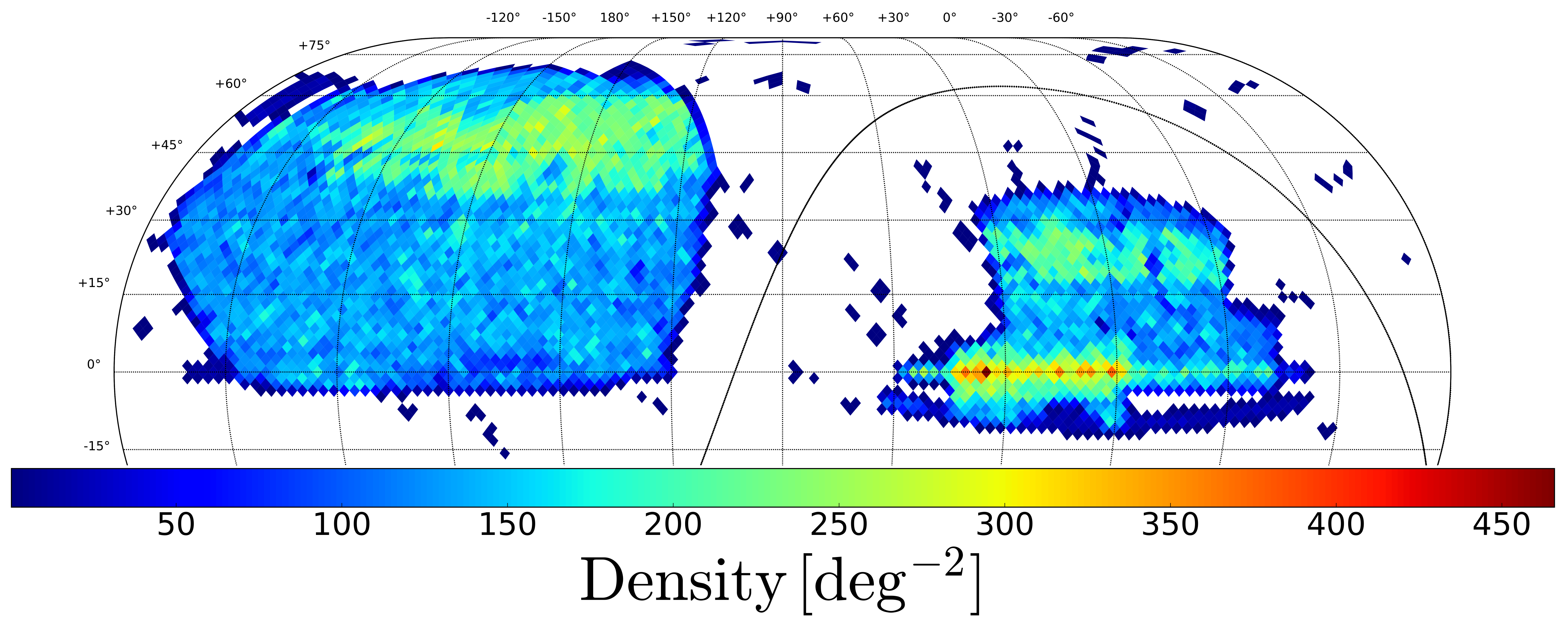}
    \caption{
    Distribution of the \num{246697} quasars and \num{1346776} galaxies used as discrete
    tracers of the matter density fluctuations. These objects have redshifts
    $0.21 < z < 1.76$. The quasars are drawn from SDSS DR14Q and the
    galaxies from SDSS DR7, BOSS DR12, and eBOSS DR14.
    }
    \label{figure::footprint}
\end{figure*}

\begin{figure}
    \centering
    \includegraphics[width=0.98\columnwidth]{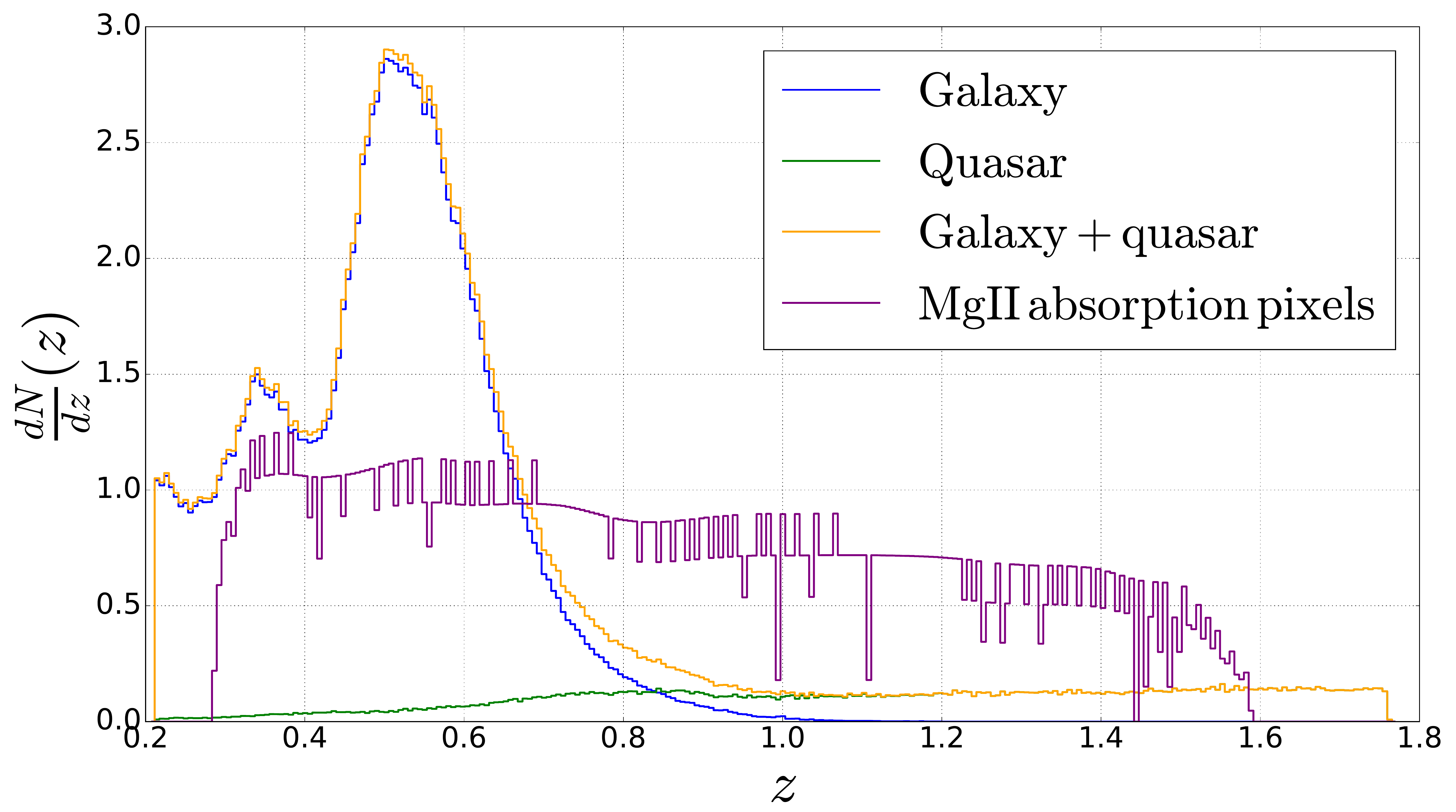}
    \caption{
    Normalized redshift distribution of pixels tracing MgII absorption (delta),
    quasars and galaxies used as tracers of the matter density fluctuations.
    The flux absorption pixels
    are given by the $\mathrm{MgII(2796)}$ transition:
    $z = \lambda_{i}/\lambda_{\mathrm{MgII(2796)}} - 1$.
    The redshift range of discrete tracers is fixed to lie within $0.21 < z < 1.76$.
    The redshift range of the pixels is bound by $0.29 < z < 1.59$.
    The pixel distribution has an apparent discretization produced by
    sky emission lines and Milky Way absorption features that are masked in this analysis.
    The quasar catalog and pixels are taken from SDSS DR14Q.
    The galaxy catalog is taken from
    SDSS DR7, BOSS DR12 and eBOSS DR14.
    }
    \label{figure::histogram_forests_tracer}
\end{figure}

In this study, we use discrete tracers with redshift $0.21 < z < 1.76$;
this range
is determined by the spectrograph efficiency, the sky emission,
the wavelength of the $\mathrm{MgII(2796)}$ absorption and the scale of BAO
(sec.~\ref{subsection::measurement_of_the_MgII_flux_fransmission_field},
sec.~\ref{subsection::correlation_function_calculation}).
Throughout, we refer to these discrete tracers as simply ``objects''.

In the redshift range relevant
to our study, $0.21 < z < 1.76$, we have \num{246697} quasar tracers observed in SDSS-I, II, BOSS and
eBOSS.
From the DR14Q catalog, we obtain the sample of quasars which are the
background to the different forests from which we measure the $\mathrm{MgII(2796)}$
absorption. We keep only objects observed in BOSS \citep{2012ApJS..199....3R} and eBOSS \citep{2015ApJS..221...27M} because
the small fraction of
DR7 data not re-observed in BOSS or eBOSS have been observed with a different spectrograph and have been
processed with a different pipeline.
We remove all objects with a broad absorption line (BAL) feature following
the automated index \texttt{BI\_CIV} in DR14Q.
Removing these peculiar objects improves the significance of our measurement.
We also remove the few quasars with $z > 5$ since their number density
is low and thus do not contribute significantly to our measurement.
This final sample of background quasars is composed of \num{500589} objects
with redshift $0.35 < z < 5$.

In the redshift range relevant
to our study, $0.21 < z < 1.76$, we have
\num{94472} eBOSS galaxies,
\num{1197675} BOSS galaxies and
\num{170151} SDSS DR7 galaxies.
We combine these three catalogs since they have similar bias that follows the
same empirical law (left panel of figure~\ref{figure::evolution_bias_qso_galaxy__evolution_bias},
described in sec.~\ref{section::Model_for_the_cross_correlations}).
We remove possible duplicates across catalogs by excluding galaxies within
one arc-second of another galaxy. In a similar manner,
we remove duplicates between the galaxy and quasar catalogs.
The final galaxy tracer catalog is composed of \num{1346776} objects.
The celestial footprint of the galaxy and quasar tracers is given in
figure~\ref{figure::footprint} and their redshift distribution,
as well as the MgII absorption pixels,
in figure~\ref{figure::histogram_forests_tracer}.

\subsection{Measurement of the flux transmission field}
\label{subsection::measurement_of_the_MgII_flux_fransmission_field}

To compute the fluctuation of flux transmission in the \num{500589} background quasars
of redshift \mbox{$0.35 < z < 5$}, we
use the Python ``Package for IGM Cosmological-Correlations Analyses''
(\texttt{picca}\footnote{\href{https://github.com/igmhub/picca}{{https://github.com/igmhub/picca}}}).
This package has been used to perform an analysis of BAO
in the cross-correlation of BOSS Ly$\alpha$ forests and quasars 
\citep[][hereafter ``dMdB2017'']{2017A&A...608A.130D}
and in the cross-correlation of eBOSS CIV forests and quasars
\citep[][hereafter ``Blomqvist2018'']{2018JCAP...05..029B}.
Using a catalog of quasars such as that described in
section~\ref{subsection::Catalog_of_quasars_and_galaxies}, \texttt{picca} processes
all of the spectra, including those with multiple epochs.
The main purpose of the package is
to compute the mean unabsorbed continuum of each quasar and
to compute the flux decrement at each pixel for each forest.
The same package also computes and fits the cross-correlation functions.

The spectra are processed using the final eBOSS pipeline \texttt{v5\_11\_0}
\citep{2012AJ....144..144B,
2017ApJS..233...25A} that will be used for DR16.
For each background DR14Q quasar, we co-add all the available good observations
from the \texttt{spPlate} files.
\begin{figure*}
    \centering
    \includegraphics[width=0.98\columnwidth]{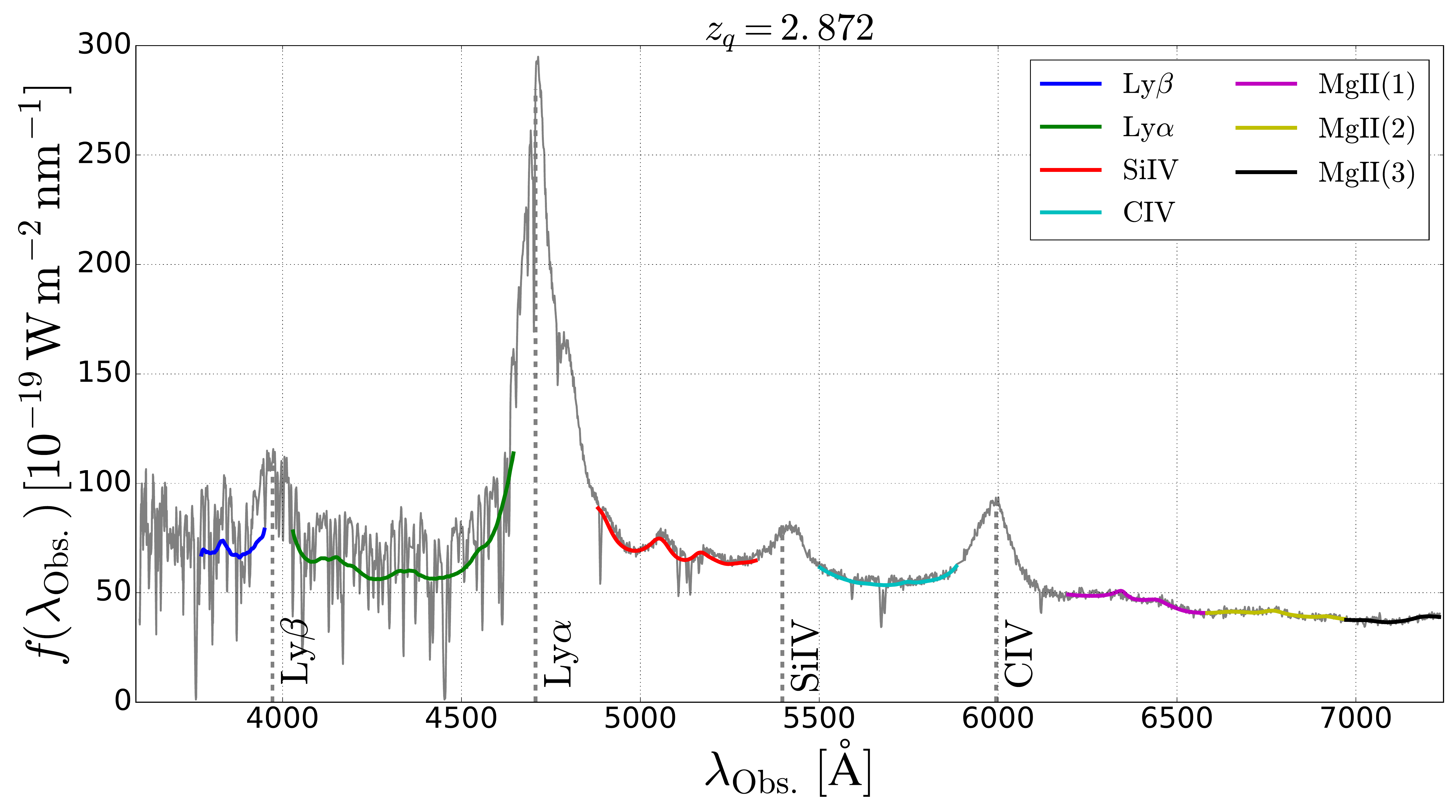}
    \includegraphics[width=0.98\columnwidth]{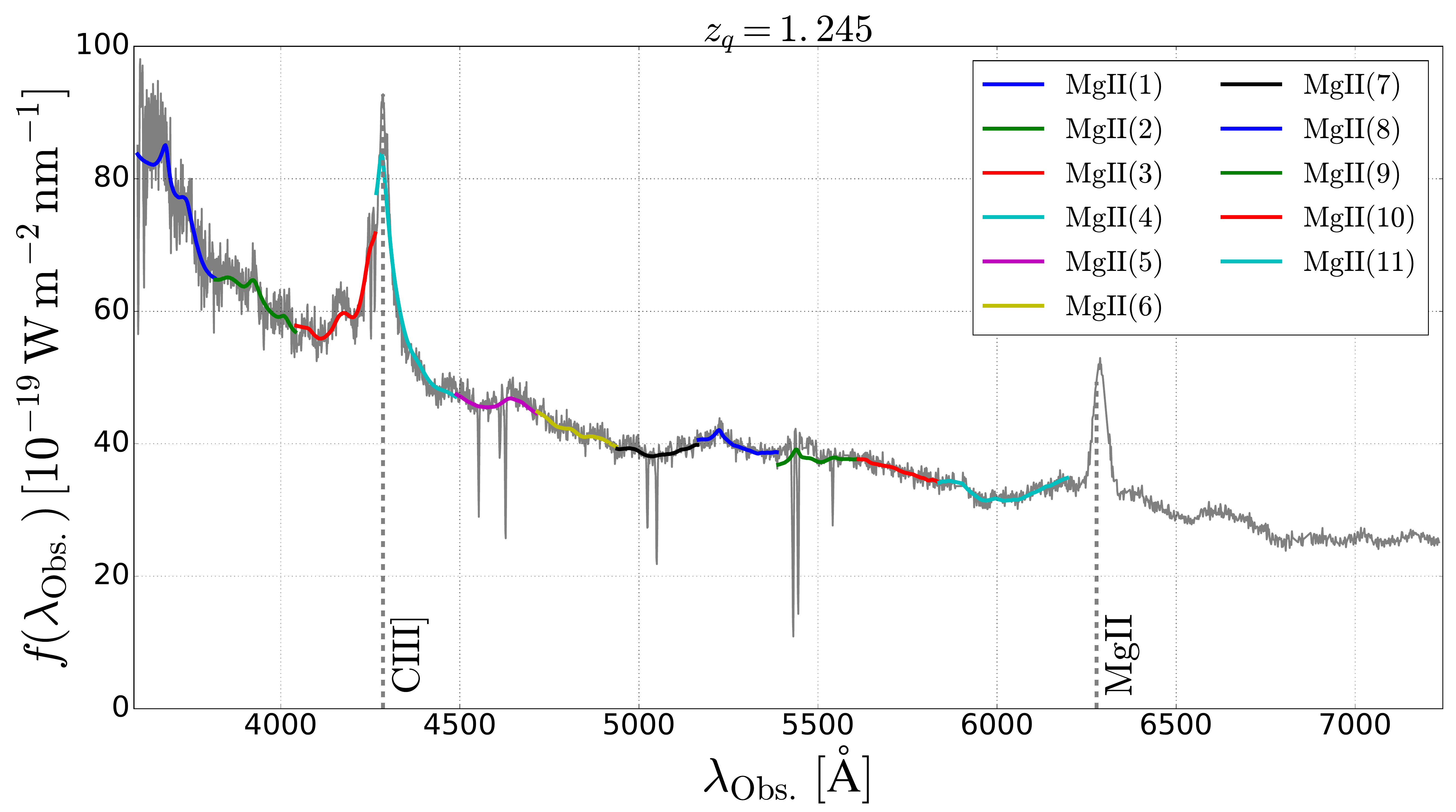}
    \caption{
    Example of two eBOSS quasars and the independent fit of all the forests of
    table~\ref{table::definition_forests}. The different colored lines indicate the quantity
    $\overline{F}(\lambda)C_{q}(\lambda_{\mathrm{R.F.}})$, from
    equation~\ref{equation::definition_delta}.
    Left: the quasar has a redshift $z = 2.872$ and is identified by
    $(\mathrm{Plate},\mathrm{MJD},\mathrm{Fiber}) = (5138,55830,20)$.
    Right: the quasar has a redshift $z = 1.245$ and is identified by
    $(\mathrm{Plate},\mathrm{MJD},\mathrm{Fiber}) = (4300,55528,224)$.
    The gray dashed lines are the location of the Ly$\beta$, Ly$\alpha$, SiIV, CIV, CIII] and MgII
    quasar emission lines.
    }
    \label{figure::exemple_data_forest}
\end{figure*}

To reduce the variance of the spectral pixels, we keep only data with
observed wavelength $\lambda \in [3600,7235] \, \text{\AA{}}$. The lower
bound of this range is set by the low system throughput at shorter wavelengths.
The upper bound is given by the increasing number of sky emission lines.
We mask small intervals of
the observed wavelength range corresponding to remaining sky emission lines
and Milky Way absorption CaII H\&K (DR14 line mask in
\href{https://github.com/igmhub/picca/blob/master/etc/dr14-line-sky-mask.txt}
{\texttt{picca}}).

As observed in multiple analyses \citep[e.g.,][]{2013A&A...552A..96B}
the eBOSS pipeline produces flux calibration errors from uncertainties
in the features of spectral standard F-star templates and sky emission.
This miscalibration results in errors at the $3\%$ level on small wavelength scales.
Furthermore, the pipeline estimates of the pixel variance are biased by up to $30\%$.
In this study, we use the flux on the red side of the MgII emission line,
$\lambda_{\mathrm{R.F.}} \in [2900,3120] \, \text{\AA{}}$, to correct
for these two aspects. This interval of the background quasar spectra is largely free from
IGM absorption, including MgII. To
compute the necessary corrections, we analyze data from the longer wavelengths
in the same manner as described below for the different forests. The correction
of the flux calibration has no systematic impact on our final measurement, and the correction
of the variance estimates only improves the significance of the final result.

To limit the loaded memory and increase the speed of the extraction of
flux transmission measurements, we combine three pipeline pixels into one analysis
pixel. The resulting width in observed wavelength is
$\Delta \log_{10}{\lambda} = 3 \times 10^{-4}$.
In the following, we refer to this combined pixel
as simply a pixel.
We also divide the spectra into $15$ different intervals
in rest-frame wavelength. We refer to each distinct interval as a forest.
Table~\ref{table::definition_forests} lists the definition
of each forest, while
figure~\ref{figure::exemple_data_forest} presents two examples of these forests
in quasar spectra.
Figure~\ref{figure::exemple_data_forest} and table~\ref{table::definition_forests} show that the MgII(3)
and MgII(4) forests have a contribution from the CIII](1909) emission line.
Variations on the strength of this line will produce correlation
between pixels of the same background quasar and an increase of variance in
our measurement. However, because the emission is uncorrelated from quasar
to quasar, it will not bias our measurement of the MgII-tracer
cross-correlation.
We limit this effect in other forests by excluding the pixels in the Ly$\beta$,
Ly$\alpha$, SiIV and CIV emission lines. In doing so, we maintain the same
definition of forests as previous studies (e.g., dMdB2017, Blomqvist2018).
\begin{table}
    \centering

    \scalebox{0.98}{
    \begin{tabular}{l r r r r r r}

    Name &
    $\lambda_{\mathrm{R.F.,\,\min}}$ &
    $\lambda_{\mathrm{R.F.,\,\max}}$ &
    $z_{\mathrm{q,\,\min}}$ &
    $z_{\mathrm{q,\,\max}}$ &
    $N_{q}$ &
    $N_{pix}$ \\

    &
    $[\text{\AA}]$ &
    $[\text{\AA}]$ &
    &
    &
    &
    $[10^{6}]$
    \\

    \noalign{\smallskip}
    \hline \hline
    \noalign{\smallskip}

    $\mathrm{Ly\beta}$  & $974$  & $1020$ & $2.65$ & $5.00$ & $\num{64041}$  & $4$  \\
    $\mathrm{Ly\alpha}$ & $1040$ & $1200$ & $2.10$ & $5.00$ & $\num{187771}$ & $31$ \\
    $\mathrm{SiIV}$     & $1260$ & $1375$ & $1.71$ & $4.55$ & $\num{246915}$ & $30$ \\
    $\mathrm{CIV}$      & $1420$ & $1520$ & $1.45$ & $3.92$ & $\num{285954}$ & $27$ \\
    $\mathrm{MgII(1)}$  & $1600$ & $1700$ & $1.19$ & $3.37$ & $\num{317005}$ & $27$ \\
    $\mathrm{MgII(2)}$  & $1700$ & $1800$ & $1.07$ & $3.11$ & $\num{319706}$ & $25$ \\
    $\mathrm{MgII(3)}$  & $1800$ & $1900$ & $0.96$ & $2.88$ & $\num{315419}$ & $23$ \\
    $\mathrm{MgII(4)}$  & $1900$ & $2000$ & $0.86$ & $2.68$ & $\num{308755}$ & $22$ \\
    $\mathrm{MgII(5)}$  & $2000$ & $2100$ & $0.77$ & $2.50$ & $\num{292241}$ & $19$ \\
    $\mathrm{MgII(6)}$  & $2100$ & $2200$ & $0.69$ & $2.33$ & $\num{263193}$ & $17$ \\
    $\mathrm{MgII(7)}$  & $2200$ & $2300$ & $0.62$ & $2.18$ & $\num{228279}$ & $14$ \\
    $\mathrm{MgII(8)}$  & $2300$ & $2400$ & $0.55$ & $2.04$ & $\num{209410}$ & $12$ \\
    $\mathrm{MgII(9)}$  & $2400$ & $2500$ & $0.49$ & $1.91$ & $\num{185835}$ & $11$ \\
    $\mathrm{MgII(10)}$ & $2500$ & $2600$ & $0.43$ & $1.80$ & $\num{163280}$ & $9$  \\
    $\mathrm{MgII(11)}$ & $2600$ & $2760$ & $0.35$ & $1.69$ & $\num{165784}$ & $13$ \\

    \end{tabular}
    }
    
    \caption{
    Definition of the $15$ different forests used in this study. The columns list 
    1) the name of the forest,
    2,3) the rest-frame wavelength range,
    4,5) the background quasar redshift range
    that have at least $50$ pixels in the given forest,
    with observed wavelength in $[3600,7235] \, \text{\AA{}}$,
    and 6,7) the total number of background quasars and pixels that contribute
    to each region.
    }
    \label{table::definition_forests}
\end{table}

In the following, we present the definition and the computation
of the fluctuation of
flux transmission for each forest. In this analysis, each forest is treated
independently.
This method is similar to the one presented in studies of Ly$\alpha$ absorption
from Ly$\alpha$ forests \citep[][dMdB2017]{2017A&A...603A..12B}
and is the same as the method presented in studies of CIV absorption in the Ly$\alpha$, SiIV and CIV
forests (Blomqvist2018).

For each background quasar $q$, and for each forest, defined in
table~\ref{table::definition_forests}, the transmitted
flux at each pixel delta$(q,\lambda)$, is:
\begin{equation}
    \delta_{q}(\lambda) =
    \frac{ f_{q}(\lambda) } {\overline{F}(\lambda)C_{q}(\lambda_{\mathrm{R.F.}})}
    - 1,
    \label{equation::definition_delta}
\end{equation}
where $\lambda$ is the observed wavelength, and $f_{q}(\lambda)$ is the
observed flux.

In dMdB2017, $\overline{F}(\lambda)$ was the mean transmitted flux
fraction between $0$ and $1$. In this study, as in Blomqvist2018, $\overline{F}(\lambda)$ is the stack
of the flux in observed wavelength, normalized so that its average over
the full wavelength range is one.
$C_{q}(\lambda_{\mathrm{R.F.}})$ is the continuum of the given forest
for the given quasar. The product $\overline{F}C_{q}$ is thus
the mean expected flux for this quasar.
To account for variability from background quasar to background quasar we define the
continuum by:
\begin{equation}
    C_{q}(\lambda_{\mathrm{R.F.}}) =
    \overline{C}(\lambda_{\mathrm{R.F.}})
    \left[
    a_{q,0}
    +
    a_{q,1} \log{\lambda_{\mathrm{R.F.}}}
    \right],
    \label{equation::definition_continuum}
\end{equation}
where $(a_{0},a_{1})_{q}$ is a set of two free parameters fitted to the
observed flux. The mean continuum $\overline{C}(\lambda_{\mathrm{R.F.}})$
is the stack of the flux over all rest-frame wavelengths and is normalized
so that its mean over each forest is equal to one.
For the Ly$\alpha$ and Ly$\beta$ forests, we correct the shape
of the continuum for the absorption of Damped Ly$\alpha$ Absorbers (DLAs)
using the automatic DR14 catalog \citep{2009A&A...505.1087N,2012A&A...547L...1N}.
We mask pixels with more than $20\%$ absorption of the flux by the DLA.
Figure~\ref{figure::exemple_data_forest} presents the quantity
$\overline{F}(\lambda)C_{q}(\lambda_{\mathrm{R.F.}})$ for each of the $15$
forests, fitted onto two background quasars.

The weight of each delta is given by:
\begin{equation}
    1/w_{q}(\lambda) =
    \eta(\lambda) \sigma^{2}_{\mathrm{noise},q}(\lambda)
    + \sigma^{2}_{\mathrm{LSS}}(\lambda)
    + \epsilon(\lambda)/\sigma^{2}_{\mathrm{noise},q}(\lambda),
    \label{equation::definition_weight}
\end{equation}
where $\sigma_{\mathrm{noise},q} = \sigma_{\mathrm{pip},q}(\lambda)/|\overline{F}(\lambda)C_{q}(\lambda_{\mathrm{R.F.}})|$.
The first term is the contribution of the measurement error;
it is taken from the pipeline error corrected by the factor $\eta(\lambda)$.
The second term is the Large Scale Structure (LSS) variance of each forest
at a given observed wavelength. It also acts as a cap for high signal-to-noise
ratio spectra.
Finally the third term is the observed
effect at large signal-to-noise ratio linked to the mismatch between the
modeled continuum and the true observed spectra.
Each of these terms are different for each forest of table~\ref{table::definition_forests}:
the resulting $\eta$ is very similar across forests, however
$\epsilon$ and $\sigma^{2}_{\mathrm{LSS}}$ differ by orders of magnitude.
As expected, the variance due to large scale structure is high in the Ly$\alpha$ and
in the Ly$\beta$ forest, of order $10^{-1}$, whereas in other forests it is less
than $5 \times 10^{-3}$. This behavior can be observed in the left panel of
figure~\ref{figure::exemple_data_forest},
where the variance of the pixels bluewards of the Ly$\alpha$ emission line
($\lambda_{\mathrm{R.F.}} = 1215.67$~\AA{}, $\lambda = 4707$~\AA{})
is larger than the one of pixels
redwards of the emission line.
Although MgII absorption is expected to be present in all forests, its
effective contribution to the observed pixel strength varies
considerably across them.

As explained in dMdB2017 and \citet{2017A&A...603A..12B}, the fit of the continuum of
equation~\ref{equation::definition_continuum} produces a distortion of the
delta field. As it was done in these previous studies, we decided to make
this distortion exact by redefining our field by:
\begin{equation}
    \delta_{q}(\lambda)
    \rightarrow
    \delta_{q}(\lambda) - \overline{\delta_{q}}
    - \left( \Lambda - \overline{\Lambda_{q}} \right)
    \frac{ \overline{ \left( \Lambda - \overline{\Lambda_{q}} \right) \delta_{q} } }
    { \overline{ \left( \Lambda - \overline{\Lambda_{q}} \right)^{2} } },
\end{equation}
where $\Lambda \equiv \log{\lambda}$, and the mean is taken over all pixels of
a given background quasar forest $q$.
The second step of making this bias exact is to subtract the mean delta
in bins of observed wavelength:
\begin{equation}
    \delta_{q}(\lambda)
    \rightarrow
    \delta_{q}(\lambda) - \overline{\delta}(\lambda).
\end{equation}

The quantities $\overline{F}$, $\overline{C}$, $\eta$,
$\sigma^{2}_{\mathrm{LSS}}$, and $\epsilon$ are computed for each of the
$15$ forests via an iterated process until they all converge.
This computation results in a total of $284\times10^{6}$ measurements of the
flux-transmission, tracing the fluctuations of MgII density in the IGM.
The statistics per forest are given in table~\ref{table::definition_forests}.
Figure~\ref{figure::histogram_forests_tracer} presents
the redshift distribution of these pixels assuming all the
absorption is from MgII.
This distribution has an apparent discretization produced by
sky emission lines and Milky Way absorption features which are masked in this analysis.

%
%
\section{The one-dimensional pixel auto-correlation}
\label{section::The_one_dimensional_pixel_auto_correlation}

This section presents the measurement of the auto-correlation of pixels
from the same forest and from the same background quasar. This correlation allows us to identify
all the different metal absorptions present in our measurement of the MgII - quasar
and MgII - galaxy cross-correlations.

The normalized one-dimensional pixel auto-correlation is given by the mean of the product of
two deltas:
\begin{equation}
    \xi^{\mathrm{1D}}_{\mathrm{Norm},\,A} = \frac{
    \sum\limits_{ \lambda_{i}/\lambda_{j} \in A } w_{i}w_{j}
    \frac{ \delta_{i}\delta_{j} }{ \sigma(\lambda_{i})\sigma(\lambda_{j}) }
    }{
    \sum\limits_{ \lambda_{i}/\lambda_{j} \in A } w_{i}w_{j}
    }.
\end{equation}
In this equation $(i,j)$ is a pair of pixels from the same forest, of
transmitted flux fractions $\delta_{i}$ and $\delta_{j}$, of weights $w_{i}$ and $w_{j}$,
and of observed wavelengths $\lambda_{i}$ and $\lambda_{j}$
(eqns.~\ref{equation::definition_delta} and \ref{equation::definition_weight}).
\mbox{$\sigma^{2}(\lambda_{i}) = \langle\delta^{2}(\lambda_{i})\rangle$} is the
measured variance of the delta at a given observed wavelength.
$A=(\lambda_{1}/\lambda_{2})_{A}$ is one bin of the correlation. The ratio
$\lambda_{1}/\lambda_{2}$ lies in the interval
$[1,\lambda_{\mathrm{R.F.,\,\max}}/\lambda_{\mathrm{R.F.,\,\min}}]$,
different for each forest.
As it is defined, the function gives the physical correlation within $[-1,1]$
for all pairs of pixels at a given separation of wavelength.
This function is exactly $1$ ($100\%$ correlated) for
$\lambda_{1} = \lambda_{2}$.

\begin{figure}
    \centering
    \includegraphics[width=0.98\columnwidth]{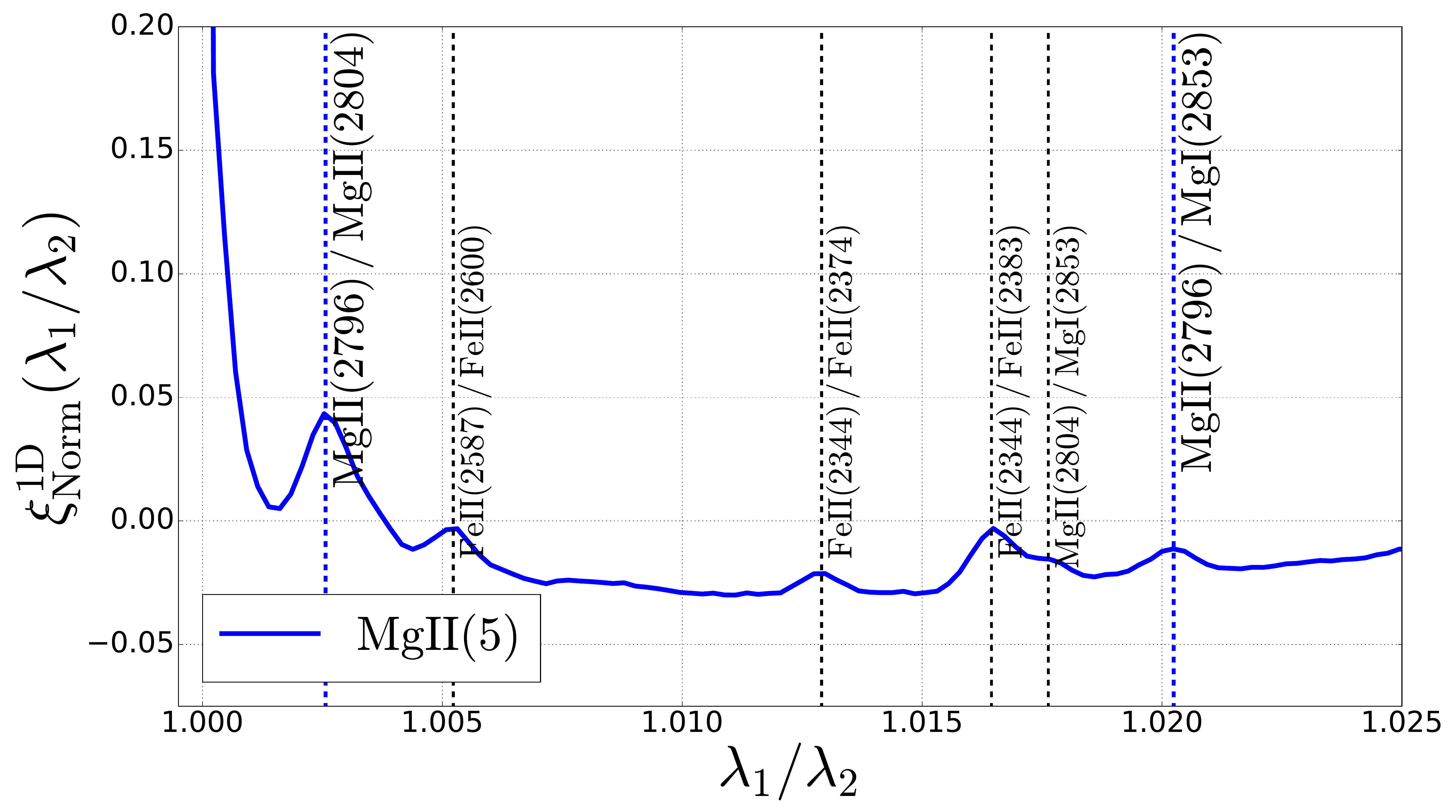}
    \caption{
    Normalized 1D auto-correlation of pixels from the same background quasar
    in the MgII(5) forest.
    The two blue dashed lines represent the
    $\mathrm{MgII(2796)}$ - metal correlation relevant to our study. The other black dashed
    lines show metal~1 - metal~2 correlations irrelevant to this study.
    This correlation was computed using SDSS \texttt{spCFrame} instead of \texttt{spPlate}
    using a pixel size of $\Delta \log_{10}{\lambda} = 1 \times 10^{-4}$.
    }
    \label{figure::cf1d_corr_MGII5_withlines}
\end{figure}

This correlation function is different for each forest defined in table~\ref{table::definition_forests}.
The auto-correlations are presented in dMdB2017 for the Ly$\alpha$ forest,
and in Blomqvist2018 for the SiIV and CIV forests.
Figure~\ref{figure::cf1d_corr_MGII5_withlines} presents
the auto-correlation only for the MgII(5) forest; other MgII(i) 1D auto-correlations are
similar.
Several peaks corresponding to flux absorbed by correlated metals are identified
in this figure.

To know which metal transitions will impact our cross-correlation study,
we must incorporate the maximal wavelength separation relevant to the scales
explored in the cosmology analysis.
To study the BAO scale, we measure the MgII - object cross-correlation up to $\pm 200 \, \hMpc$
along the line-of-sight (sec.~\ref{subsection::correlation_function_calculation}).
At $\lambda = 4600 \, \text{\AA}$,
this distance translates into a wavelength ratio of $\lambda_{1}/\lambda_{2} = 1.06$,
with respect to $\mathrm{MgII(2796)}$.
We use the measured stack of absorption
in quasar spectra from \citet{2006MNRAS.367..945Y}, \citet{2014MNRAS.441.1718P} and
\citet{2017ApJ...846....4M} to list all the metal transitions such that
$\lambda_{1}/\lambda_{\mathrm{MgII(2796)}} \in [1/1.06,1.06]$.
Only three metal transitions satisfy this condition:
MgII(2796), our reference,
MgII(2804), the other doublet member,
and MgI(2853), absorption from neutral magnesium. Information on these two transitions is provided in
table~\ref{table::metals_contribution}.
The presence of these three metals in our quasar spectra data set is confirmed by the two correlation
peaks marked by dashed blue lines in figure~\ref{figure::cf1d_corr_MGII5_withlines}:
$\mathrm{MgII(2796) \, / \, MgII(2804)}$ and $\mathrm{MgII(2796) \, / \, MgI(2853)}$.
\begin{table}
    \centering
    \scalebox{1.0}{
    \begin{tabular}{l l l l c }
        $\mathrm{Transition}$ &
        $\lambda_{\mathrm{R.F.}}$ &
        $\lambda_{1}/\lambda_{\mathrm{MgII(2796)}}$ &
        $r_{\parallel}$ \\

        &
        $[\text{\AA}]$ &
        &
        $[\hMpc]$ \\

        \noalign{\smallskip}
        \hline \hline
        \noalign{\smallskip}
        MgII(2796) & $2796.35$ & $1$ & $0$ \\
        MgII(2804) & $2803.53$ & $1.0026$ & $+9$ \\
        MgI(2853)  & $2852.96$   & $1.0202$ & $+68$ \\
        \noalign{\smallskip}
        \noalign{\smallskip}
    \end{tabular}
    }
    \caption{
    List of the metal transitions present in the
    MgII - object cross-correlation. The columns are
    1) the name of the transition,
    2) the rest-frame wavelength,
    3) the ratio $\lambda_{1}/\lambda_{\mathrm{MgII(2796)}}$, and
    4) the expected shift in $\hMpc$, according to the $\Lambda$CDM cosmology,
    of the cross-correlation
    at $\lambda = 4446 \, \text{\AA}$ ($z=0.59$ for \mbox{MgII(2796)}).
    }
    \label{table::metals_contribution}    
\end{table}

Figure~\ref{figure::cf1d_corr_MGII5_withlines} displays other metal correlations
marked by black dashed lines involving different FeII transitions,
but not involving MgII(2796).
These metal transitions produce peaks in our correlation that are too far
from our separation of $\pm 200 \, \hMpc$ along the line-of-sight. Thus, they
are irrelevant to our cross-correlation study.
Contrary to pixel auto-correlations, a pixel-object cross-correlation
can only mis-interpret the redshift of the pixel because
the redshift of the object is measured with low catastrophic failure rate.
The consequence is that when using MgII(2796)
as the reference redshift, absorption due to MgII(2796) will produce a peak
in the cross-correlation at $r_\parallel{} = 0 \, \hMpc{}$,
while absorption due to MgI(2853) will produce a peak in the cross-correlation at
$r_\parallel{} \sim +68 \, \hMpc{}$ (table~\ref{table::metals_contribution})
given the effective redshift of our sample.
For FeII(2600), the peak is at $-248 \, \hMpc{}$; the other FeII transitions
are even further remote from our $\pm 200 \, \hMpc$ range.

%
%
\section{The MgII - quasar and MgII - galaxy cross-correlation}
\label{section::The_MgII_quasar_and_MgII_galaxy_cross_correlation}

This section presents the measurement of the MgII - quasar and 
MgII - galaxy cross-correlation in each forest, along with
their associated covariance matrices and the model to account for
distortions introduced by continuum fitting.

\subsection{The correlation function}
\label{subsection::correlation_function_calculation}

The biased cosmological cross-correlation is calculated independently
for each set of forest-object pairs. We follow the same
techniques as in \citet{2012JCAP...11..059F,2013JCAP...05..018F} and dMdB2017.
The cross-correlation is given by the weighted mean of delta from one forest at a given separation of an object:
\begin{equation}
    \xi^{q\,f}_{A} = \frac{\sum\limits_{(i,k) \in A} w_{i} \, \delta_{i} }{
    \sum\limits_{(i,k) \in A} w_{i} }.
    \label{equation::cross_correlation_definition}
\end{equation}
In this equation $i$ is a pixel of one of the forests
(table~\ref{table::definition_forests}) of transmitted flux fraction $\delta_{i}$
and weight $w_{i}$.
The sum runs over all possible pixel $(i)$ - object $(k)$ pairs
falling inside the bin $A = \left(r_{\parallel},r_{\perp}\right)_{A}$.
We reject pairs involving a quasar and a pixel from its own forest,
since the correlation vanishes for these pairs
due to the fit of the continuum of equation~\ref{equation::definition_continuum}.

The distance along the line-of-sight, or parallel distance, $r_{\parallel}$, and the
distance across the line-of-sight, or perpendicular distance, $r_{\perp}$, are given by:
\begin{equation}
    \begin{array}{ll}
        r_{\parallel} = \left[ D_{M}(z_{i})-D_{M}(z_{k}) \right]
        \cos
        \left( \frac{\Delta \theta}{2} \right),
        \\
        r_{\perp} = \left[ D_{M}(z_{i})+D_{M}(z_{k}) \right]
        \sin
        \left( \frac{\Delta \theta}{2} \right).
    \end{array}
    \label{equation::definition_distances}
\end{equation}
In this study we will also use the quantity $\vec{r} = (r,\mu)$,
where $r^{2} = r_{\parallel}^{2} + r_{\perp}^{2}$ and $\mu = r_{\parallel}/r$.
In the two relationships defined in equation~\ref{equation::definition_distances},
$\Delta \theta$ is the angle between the pixel in the forest and the object
on the celestial sphere, $z_{k}$ is the redshift of the object, and
$z_{i} = \lambda_{i}/\lambda_{\mathrm{MgII(2796)}} - 1$ is the redshift of the pixel
assuming the absorption is due to the metal transition $\mathrm{MgII(2796)}$.
Finally, the comoving angular distance, $D_{M}(z)$, is computed assuming the fiducial
flat-$\Lambda$CDM cosmology of \citet{2016A&A...594A..13P} (TT+lowP combination):
\begin{equation}
    \begin{array}{ll}
        \Omega_{\rm c}h^{2} = 0.1197,\,
        \Omega_{\rm b}h^{2} = 0.02222,\,
        \Omega_{\nu}h^{2} = 0.0006,
        \\
        h = 0.6731,\,
        N_{\nu} = 3,\,
        \sigma_{8} = 0.830,\,
        n_{s} = 0.9655.
    \end{array}
\end{equation}
This cosmology has a density of matter $\Omega_{m} = 0.315$,
a density of dark energy $\Omega_{\Lambda} = 1-\Omega_{m} = 0.685$,
a growth rate of structure $f(z=0.59) = 0.79$, and $\sigma_{8}(z=0.59) = 0.61$.

Given the fiducial cosmology, we compute the sound horizon at the drag epoch
using CAMB \citep{2000ApJ...538..473L}: $r_{d} = 99.17 \, \hMpc$.
To correctly study
the BAO scale, we compute the cross-correlation of
equation~\ref{equation::cross_correlation_definition}
to approximately twice the BAO scale, $2r_{d} \sim 200 \, \hMpc$, in both directions.
We thus limit the computation
to $r_{\parallel} \in [-200,200] \, \hMpc$ and to $r_{\perp} \in [0,200] \, \hMpc$,
with a bin size of $4 \, \hMpc$ in both dimensions. With these selections,
the correlation function
has $N_{\mathrm{bin}} = 100 \times 50 = 5000$ bins.

The observed wavelength coverage of the pixels is
$\lambda_{i} \in [3600,7235] \, \text{\AA}$
(sec.~\ref{subsection::measurement_of_the_MgII_flux_fransmission_field}).
Given the definition of the redshift from the $\mathrm{MgII(2796)}$ absorption,
the redshift range covered by the pixels is $z_{i} \in [0.29,1.59]$.
Since we compute the correlation for
$r_{\parallel} \in [-200,200] \, \hMpc$ along the line-of-sight, any objects
with redshift $z_{k} \in [0.21,1.76]$ can potentially be in a pixel-object pair
according to our $\Lambda$CDM cosmology.
We reduce the computation time by removing any object outside of this
interval.

\begin{figure}
    \centering
    \includegraphics[width=0.98\columnwidth]{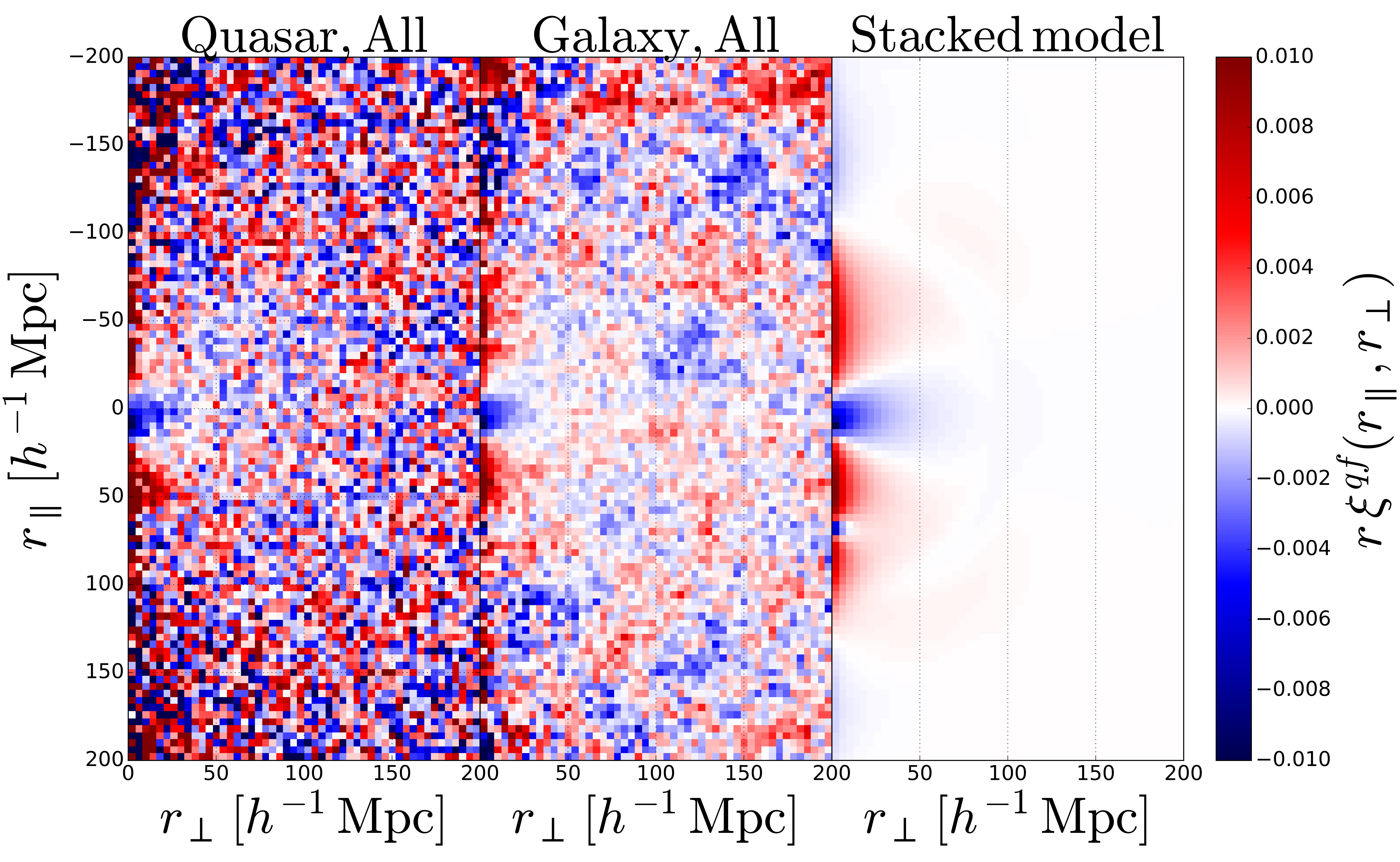}
    \caption{
    Measured and best fit MgII - object cross-correlation multiplied
    by the absolute separation $r = |\vec{r}|$.
    The color scale is saturated at both negative and positive values
    in order not to be dominated by the noise and to
    be the same for the three figures; it is symmetric about zero.
    Left: Stack of the $15$ correlations involving quasars.
    Center: Stack of the $15$ correlations involving galaxies.
    Right: Stack of the $30$ best fit models, when running the combined
    fit (last line of table~\ref{table::result_fit}).
    }
    \label{figure::corr_2d_all_quasar__all_galaxy__cutted}
\end{figure}

We compute the cross-correlation of equation~\ref{equation::cross_correlation_definition}
for all the different forest ($f$) and object ($q$) pairs. We have $15$ forests
(table~\ref{table::definition_forests}) and two objects
(quasar or galaxy), yielding a measurement of $30$ different correlation functions.
Figure~\ref{figure::corr_2d_all_quasar__all_galaxy__cutted} presents the stack of the
$15$ quasar - forest cross-correlations
on the left and the stack of the $15$ galaxy - forest cross-correlations in the center.
Both correlation functions are multiplied by the absolute
separation $r = |\vec{r}|$ for illustrative purposes.
The color scale is saturated at both negative and positive values
in order not to be dominated by the noise; it is symmetric about zero.
Both correlations are negative at small separations, indicating an increased
probability of having absorption by MgII when the pixel is near an object.
Both figures present at $r_{\parallel} \in [-50,50] \, \hMpc$ and $r_{\perp} \sim 0 \, \hMpc$
a succession of positive correlation (red),
zero correlation (white), negative correlation (blue),
then back to zero and positive correlation. This is the mark of two effects
on the cross-correlations: redshift-space distortions from the velocity of
MgII and quasars, and the effect of the distortion
matrix described in section~\ref{subsection::The_distortion_matrix}.
In particular, the distortion
of the correlation function along the radial direction can lead to a
change of sign in the amplitude of the clustering, as indicated in the
model shown in the right panel of figure~\ref{figure::corr_2d_all_quasar__all_galaxy__cutted}.
Because of the level of noise, the BAO scale can not be seen in such
figures.

For each of our $30$ cross-correlations we define the effective redshift,
$z_{\mathrm{eff}}$, as the weighted mean redshift of object-pixel pairs for bins
with $r \in [80, 120] \, h^{-1} \, \mathrm{Mpc}$, i.e., in the region where
the BAO feature is expected according to our $\Lambda$CDM cosmology.
The effective redshift values are given in
table~\ref{table::result_fit}; they range from $z_{\mathrm{eff}} = 0.44$
to $z_{\mathrm{eff}} = 1.03$, with an effective redshift for the weighted
stack of all cross-correlations: $z_{\mathrm{eff}} = 0.59$.

The number of object-pixel pairs in the BAO region,
$r \in [80, 120] \, h^{-1} \, \mathrm{Mpc}$, varies from
$134$~million in the cross-correlation between the Ly$\beta$ forest and quasars up to
$36$~billion in the cross-correlation between the Ly$\alpha$ forest and galaxies.
Over the $30$~different correlation functions, a total of
$170$~billion object-pixel pairs are used in the region where BAO is expected.

\subsection{The covariance matrix}
\label{subsection::The_covariance_matrix}

The covariance matrix of the cross-correlation is calculated by sub-sampling
the data sample similar to the approach in dMdB2017.
We divide the sky into HEALPix pixels \citep{2005ApJ...622..759G} and
compute the cross-correlation function in each
sub-sample. Using a division of the footprint of figure~\ref{figure::footprint} with 
$\textit{nside}=32$, we obtain a minimum of $3219$ and a maximum of $3266$ sub-samples
for each cross-correlation.
Each cross-correlation has $5000$ bins in $(r_{\parallel},r_{\perp})$;
the covariance between two of these bins $A$ and $B$
is given by:
\begin{equation}
    C_{AB} = \frac{1}{W_{A} W_{B}} \sum\limits_{s} W_{A}^{s} W_{B}^{s} 
    \left[ \xi^{s}_{A} \xi^{s}_{B}
    - \xi_{A} \xi_{B} \right],
    \label{equation::covar_xi_estimator_subsampling}
\end{equation}
where $\xi_{A}^{s}$ and $W_{A}^{s}$ are the cross-correlation and the sum of weights
in the sub-sample $s$, respectively, for the bin $A$.

This covariance matrix can be decomposed into two quantities.
The diagonal, $C_{AA}$, gives the variance of the bins,
and is approximately inversely proportional to the number of pairs
and proportional to the variance of the pixel:
\begin{equation}
    C_{AA} \sim \frac{c_{A} \left<\delta^{2}\right>}{N^{A}_{\mathrm{pair}}}.
\end{equation}
The variance of the pixels $\langle\delta^{2}\rangle$ is of order $0.003$
in all MgII(i) forests and higher in other forests:
of order $0.1$ in Ly$\beta$ and Ly$\alpha$ and of order $0.01$ in SiIV and CIV.
The parameter $c_{A}$ is the strength of the correlations between
different object-pixel pairs.
If all pairs are independent, $c_{A}=1$.
Since the same pixel is used in different pixel-object pairs, and since pixels
are correlated along their line-of-sights, $c_{A}$ is larger than one.
The parameter $c_{A}$ is different for each forest - object pair:
it is approximately $5$ for correlations involving MgII(i) forests and slightly
lower in other forests.

To describe the off-diagonal terms of the covariance matrix,
it is convenient to define the correlation matrix:
\begin{equation}
    Corr_{AB} = \frac{C_{AB}}{ \sqrt{C_{AA}C_{BB}} }.
    \label{equation::correlation_matrix}
\end{equation}

\begin{figure*}
    \centering
    \includegraphics[width=0.98\columnwidth]{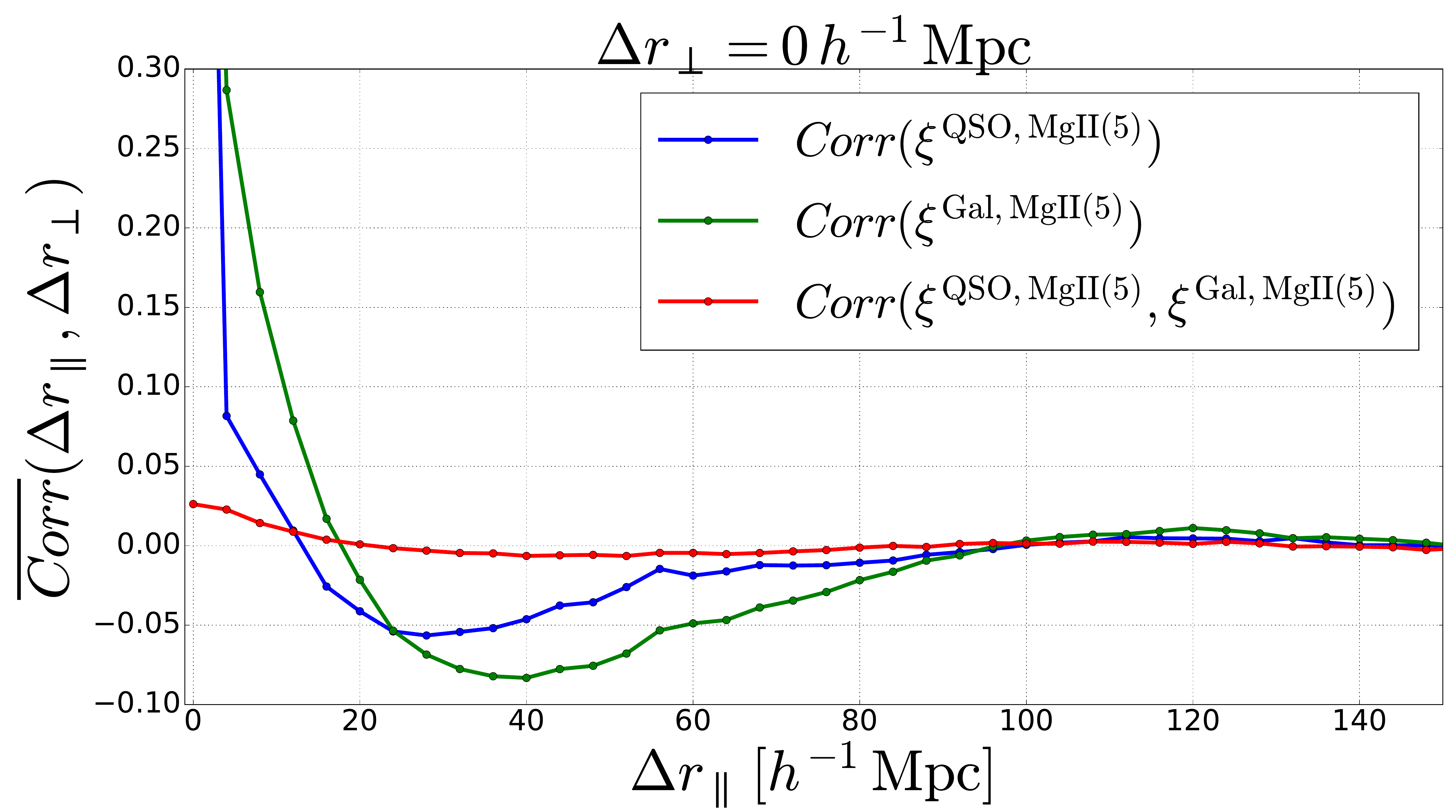}
    \includegraphics[width=0.98\columnwidth]{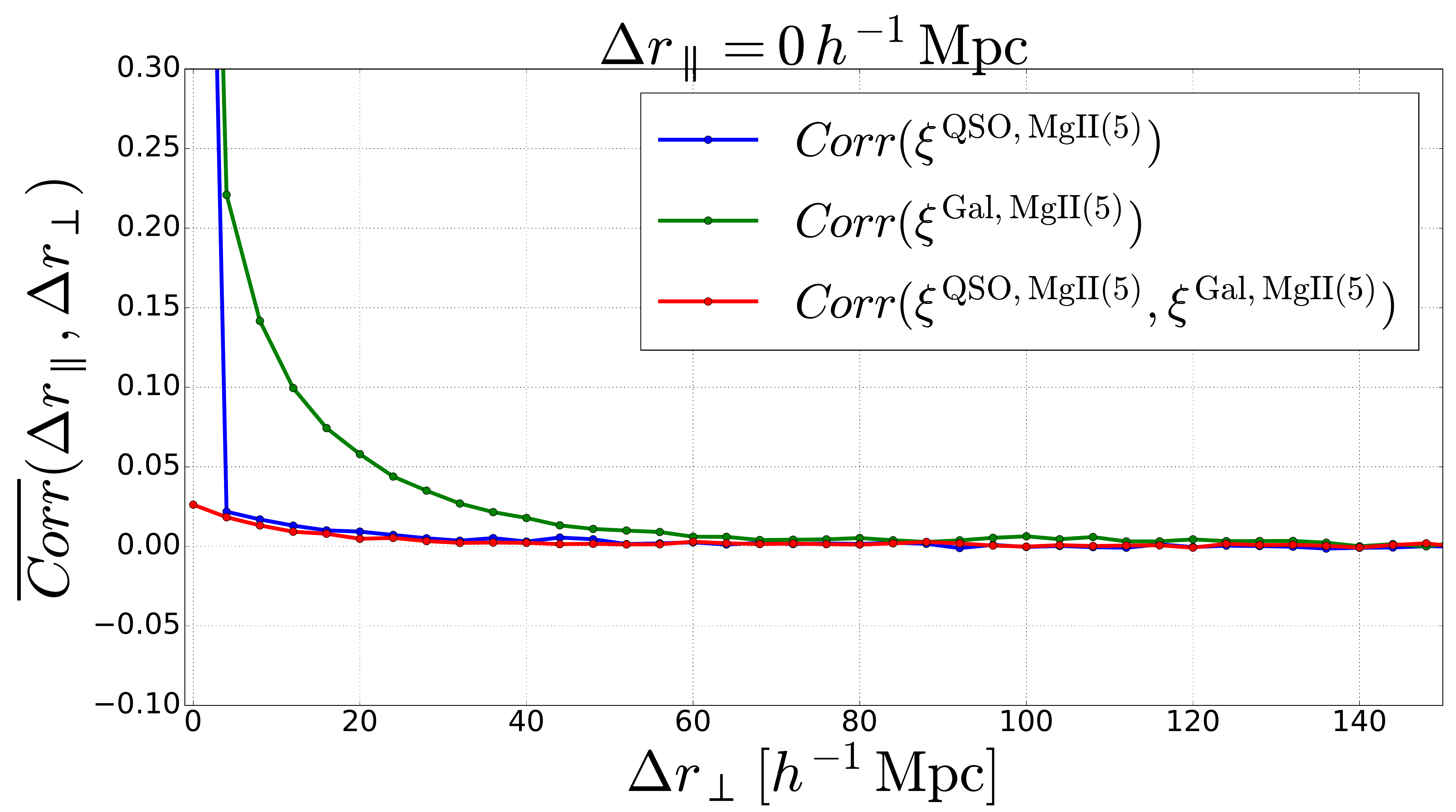}
    \caption{
    Correlation matrix $Corr_{AB}$, i.e., the normalized covariance matrix,
    for the cross-correlation $\xi^{\mathrm{QSO},\mathrm{MgII(5)}}$ (in blue)
    and for the cross-correlation $\xi^{\mathrm{Gal},\mathrm{MgII(5)}}$ (in green).
    The red curves are the cross-correlation matrix: the correlation matrix
    between the two previous cross-correlations.
    All correlations are given for a mean over all possible pairs within
    $\Delta r_{\perp} = 0 \, \hMpc$ (left) and $\Delta r_{\parallel} = 0 \, \hMpc$
    (right) as a function of $\Delta r_{\parallel}$ (left) and $\Delta r_{\perp}$ (right).
    All correlation matrices vanish for $\Delta r_{\parallel}>150 \, \hMpc$.
    }
    \label{figure::covariance_matrix_01}
\end{figure*}

As with the variance, this correlation matrix is different for each forest - object
measurement.
Figure~\ref{figure::covariance_matrix_01} displays the primary elements of the correlation matrix for the
cross-correlation between quasars and MgII absorption in the MgII(5) forest:
$\xi^{\mathrm{QSO},\mathrm{MgII(5)}}$, and for the
cross-correlation between galaxies and MgII absorption in the MgII(5) forest:
$\xi^{\mathrm{Gal},\mathrm{MgII(5)}}$.
The left panel of the figure shows the correlation matrix as a function of
$\Delta r_{\parallel} = |r_{\parallel,A}-r_{\parallel,B}|$ for a constant
$\Delta r_{\perp} = |r_{\perp,A}-r_{\perp,B}| = 0 \, \hMpc$.
The right panel of the figure presents the correlation in the other
direction: as a function of $\Delta r_{\perp}$
at $\Delta r_{\parallel} =  0 \, \hMpc$.
Both correlations drop with increasing separation; however, the correlations decrease faster
for quasars than for galaxies.
The correlations in the galaxy-forest measurement explain the different patches of
the middle panel of figure~\ref{figure::corr_2d_all_quasar__all_galaxy__cutted}.
This slower decrease is explained by the fact that there are around five times as many galaxies
as there are quasars, and thus the same pixel is used for an object-pixel pair
many more times with galaxies than with quasars.

To limit the noise due to the finite number of sub-samples, we model all the different
correlation matrices by taking their mean as a function of
$(\Delta r_{\parallel}, \Delta r_{\perp}) = (|r_{\parallel,A}-r_{\parallel,B}|, |r_{\perp,A}-r_{\perp,B}|)$.
This model was validated for the Ly$\alpha$ - quasar cross-correlation in
dMdB2017 using different methods of estimating the correlation matrix.
Three of the $30$ cross-correlations still have a non-positive definite
correlation matrix: galaxy-MgII(8), quasar-MgII(8) and galaxy-MgII(10).
We maintain their variance estimates but replace their correlation matrix with the one
from the neighboring forest: galaxy-MgII(7), quasar-MgII(7) and
galaxy-MgII(9), respectively.


Figure~\ref{figure::covariance_matrix_01} also presents the cross-correlation
matrix between the cross-correlation $\xi^{\mathrm{QSO},\mathrm{MgII(5)}}$ and
$\xi^{\mathrm{Gal},\mathrm{MgII(5)}}$, defined by:
\begin{equation}
    Corr^{q1\,f1,q2\,f2}_{AB} =
    \frac{
    C^{q1\,f1,q2\,f2}_{AB}
    }{
    \sqrt{C^{q1\,f1}_{AA}C^{q2\,f2}_{BB}}
    },
\end{equation}
where $A$ is a bin of the cross-correlation $\xi^{q1\,f1}$, of covariance
$C^{q1\,f1}$, and $B$ is a bin of the cross-correlation $\xi^{q2\,f2}$,
of covariance $C^{q2\,f2}$. The cross-covariance is given by:
\begin{equation}
    C_{AB}^{q1\,f1,q2\,f2} = \frac{1}{W_{A} W_{B}} \sum\limits_{s} W_{A}^{s} W_{B}^{s} 
    \left[ \xi^{s,\,q1\,f1}_{A} \xi^{s,\,q2\,f2}_{B}
    - \xi^{q1\,f1}_{A} \xi^{q2\,f2}_{B} \right].
\end{equation}
Since there are $15$ different forests and $2$ different objects,
we have $30$ different cross-correlations and $435$ different
cross-covariances. As seen in figure~\ref{figure::covariance_matrix_01},
the correlation between the two cross-correlations doesn't exceed $4\%$
and vanishes at large separations.
The amplitude is similar for all $435$ cross-covariances. In this study, we
thus neglect this correlation.

\subsection{The distortion matrix}
\label{subsection::The_distortion_matrix}

The fit of the continuum of equation~\ref{equation::definition_continuum}
introduces correlations between pixels from the same forest.
The larger the wavelength coverage of a forest, the smaller the correlation.
This aspect introduces extra correlation in the 3D pixel-object
cross-correlation at large scale. The measured correlation is thus a
``distorted'' version of the true cross-correlation.
The process to compute the distortion matrix, $D_{AA\prime}$, that describes the 
transformation of the true
cross-correlation to the measured cross-correlation is presented in
section~4.2 of dMdB2017.

The distortion matrix depends on the length of the forest,
the geometry of the survey, and the relative weights of the pixels.
We thus compute this matrix for each of the $30$ forest-object pairs.
The different \mbox{MgII(i)} forests are shorter than the Ly$\alpha$ forest so
their distortion matrix is less diagonal, i.e., the correlation between
pixels from the same forest is stronger.
For the Ly$\alpha$ forest, the diagonal terms cover the range $0.97 < D_{AA\prime}<0.98$ and the off-diagonal terms are
$|D_{AA\prime}|<0.022$.
For the \mbox{MgII(i)} forests, the diagonal terms cover the range $0.92 < D_{AA\prime}<0.96$ and the off-diagonal terms are
$|D_{AA\prime}|<0.077$.

%
%
\section{Fit for cosmological correlations}
\label{section::Fit_for_cosmological_correlations}

\subsection{Model for the cross-correlations}
\label{section::Model_for_the_cross_correlations}

The model fitting technique used to analyze the 30 different cross-correlations is the same as the one
developed in section~5.1 of dMdB2017
and in section~6 of Blomqvist2018.
We only give here a brief summary.

Each measured cross-correlation is a combination of three correlations.
For correlations involving quasars, they are
the quasar-\mbox{MgII(2796)}, the quasar-\mbox{MgII(2804)}
and the quasar-\mbox{MgI(2853)} cross-correlations. In this analysis,
we have defined the redshift of the pixels using the wavelength of the
\mbox{MgII(2796)} absorption. The main effect of this operation is to shift the different
correlations involving other transitions mostly along the $r_{\parallel}$
direction.
These shifts are given by the fiducial cosmology and evolve with redshift.
At a redshift $z = 0.59$, corresponding to $\lambda = 4446 \, \mathrm{\AA{}}$
for \mbox{MgII(2796)}, the shifts are $\sim9 \, \hMpc{}$ for quasar-\mbox{MgII(2804)}
and 
$\sim68 \, \hMpc{}$ for quasar-\mbox{MgI(2853)}, as given in
table~\ref{table::metals_contribution}.
This effect is taken into account by the ``metal distortion matrix''
(eqn.~6.18 of Blomqvist2018).

The expected measured signal for each of the 30 object-forest cross-correlations is given by:
\begin{multline}
    \widehat{\xi_{A}} = D_{A A^{\prime}}
    \biggl[
    \xi_{A^{\prime}}^{q,\mathrm{MgII(2796)}}\\
    + M_{A^{\prime}B^{\prime}}^{q,\mathrm{MgII(2804)}} \xi_{B^{\prime}}^{q,\mathrm{MgII(2804)}}
    + M_{A^{\prime}B^{\prime}}^{q,\mathrm{MgI(2853)}} \xi_{B^{\prime}}^{q,\mathrm{MgI(2853)}}
    \biggr],
    \label{equation::definition_expected_measur}
\end{multline}
where the sum is implicit over the repeated bins in $(r_{\parallel},r_{\perp})$,
$A^{\prime}$ and $B^{\prime}$, from equation~\ref{equation::cross_correlation_definition}.
In this equation, $q$ represents one of the two discrete tracers:
$q \in \{\mathrm{quasar}$, $\mathrm{galaxy}\}$,
$D_{A A^{\prime}}$ is the distortion matrix that models
the modification of the correlation-function by the fit of the continuum
from equation~\ref{equation::definition_continuum} (sec.~\ref{subsection::The_distortion_matrix}),
and $M_{A^{\prime}B^{\prime}}$
is the metal distortion matrix, introduced above.

Each of the three cross-correlation $\xi^{q,m}$ of
equation~\ref{equation::definition_expected_measur}
is given by the Fourier transform of the cross-power spectrum:
\begin{equation}
    P^{q,m}(\vec{k},z) \propto
    b_{q}(z)  b_{m}(z)
    \left( 1+\beta_{q}\mu_{k}^{2} \right)
    \left( 1+\beta_{m}\mu_{k}^{2} \right)
    P_{QL}(\vec{k},z,\alpha_{\mathrm{iso}}),
    \label{equation::power_spectrum}
\end{equation}
with \mbox{$m \in \{$MgII(2796)}, \mbox{MgII(2804)}, \mbox{MgI(2853)$\}$}
and with \mbox{$q \in \{$quasar}, \mbox{galaxy$\}$}.

The bias $b_{i}$ and the redshift space distortion (RSD) parameter $\beta_{i}$
are different for each tracer: \mbox{$i \in \{$quasar}, galaxy, \mbox{MgII(2796)},
\mbox{MgII(2804)}, \mbox{MgI(2853)$\}$} and evolve with redshift.
In this analysis, the three magnesium transitions are treated as a
continuum field of absorption; this implies that their bias and RSD parameters
are given, following \citet{2003ApJ...585...34M,2012JCAP...07..028F,2018MNRAS.480..610G}, by:
\begin{equation}
    b_{m}(z) = - \tau_{m}(z) b_{h,m}(z),
    \label{equation::def_transition_bias}
\end{equation}
and
\begin{equation}
    \beta_{m}(z) = \beta_{h,m}(z).
    \label{equation::def_transition_beta}
\end{equation}
In these two equations, $(b,\beta)_{m}$ are the bias and RSD parameters
of the three different magnesium transitions treated as a continuous field,
while $(b,\beta)_{h,m}$ are their respective host halos bias and RSD parameter.
The averaged optical depth, $\tau_{m}$, is different for each of the three transitions
and can evolve with redshift.
As explained in the introduction (sec.~\ref{section::Introduction}), MgII has a bias much smaller than unity
when treated as a transmission field.  This is apparent in equation~\ref{equation::def_transition_bias}.
While the halos that host MgII absorption have a bias comparable to that
of galaxies, $|b_{h,m}| \sim 1$, the mean optical depth is much smaller than unity:
$\tau_{m} \ll 1$.  Another way to understand the low bias is to compare the MgII
transmission field to that of the Lyman-$\alpha$ forest.  The cosmic MgII
number density is much lower than that of neutral hydrogen as can be seen
in the two spectra of figure~\ref{figure::exemple_data_forest}.
Many more absorption lines from hydrogen
are visible at wavelengths shorter than the quasar Ly$\alpha$ emission line than
from any other metals at wavelengths longer than the quasar Ly$\alpha$ emission line.
The bias from metals is therefore much lower than that of Lyman-$\alpha$.

\begin{figure*}
    \centering
    \includegraphics[width=0.98\columnwidth]{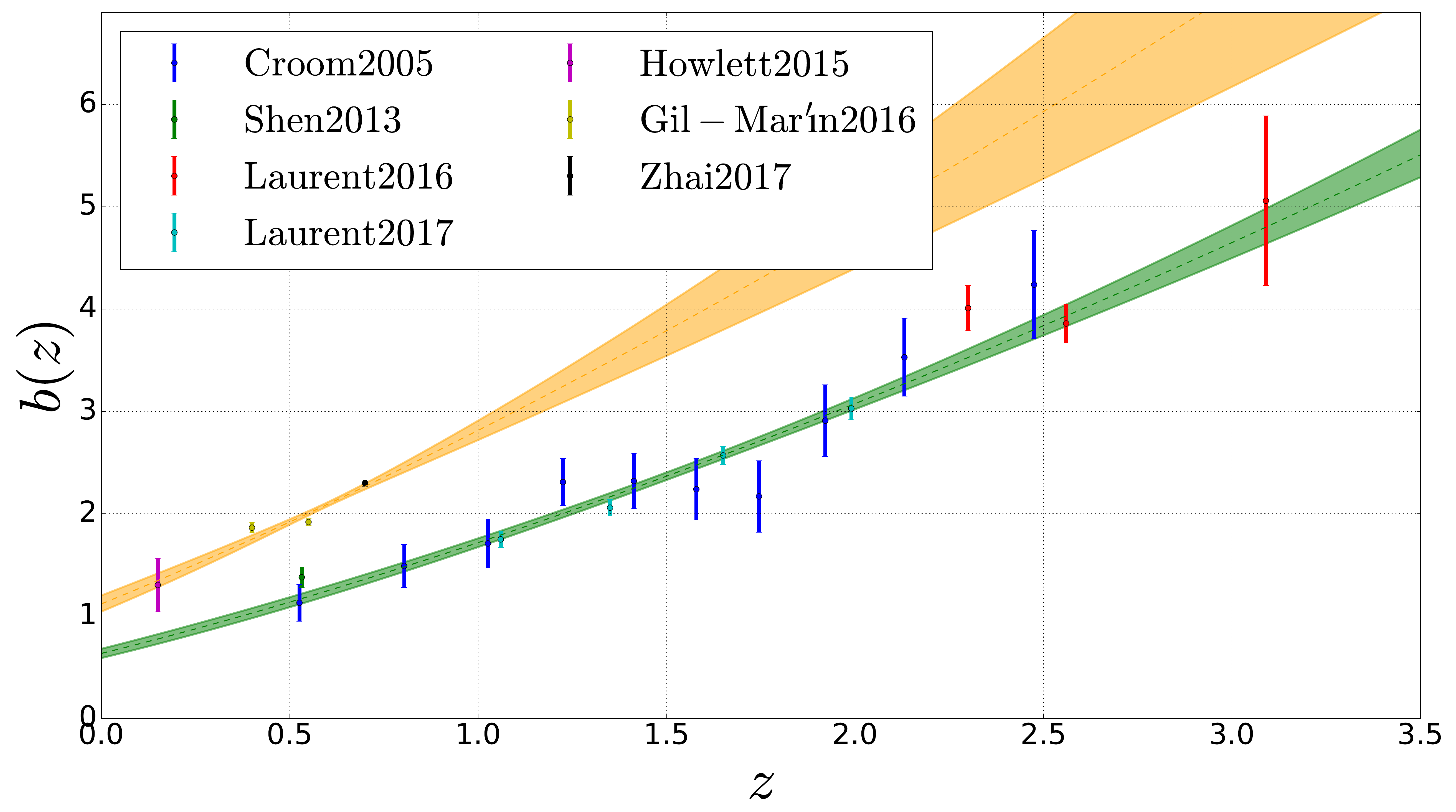}
    \includegraphics[width=0.98\columnwidth]{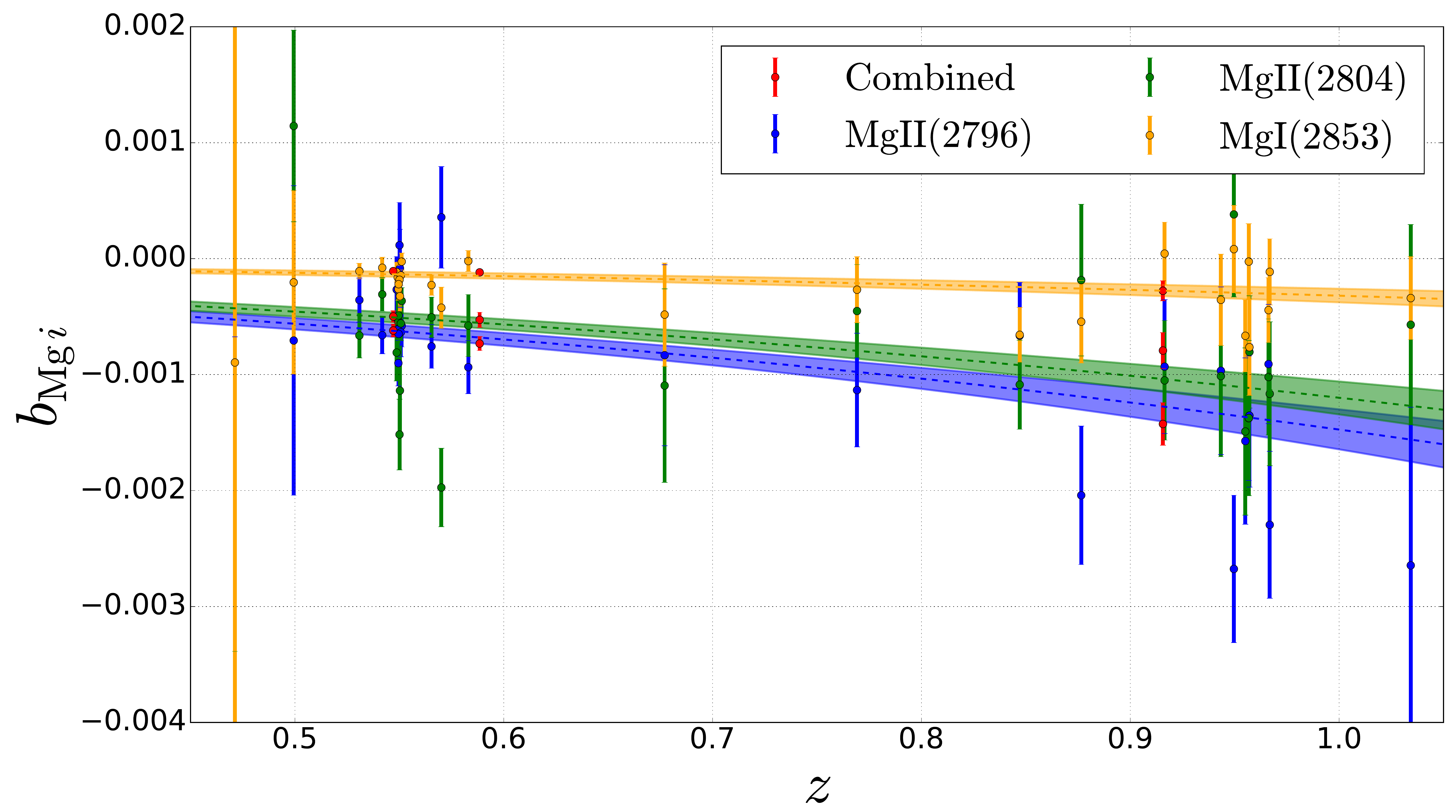}
    \caption{
    Bias evolution with redshift.
    Left: evolution for quasar bias from
    \citet{2005MNRAS.356..415C, 2013ApJ...778...98S, 2016JCAP...11..060L, 2017JCAP...07..017L},
    and for galaxies from \citet{2015MNRAS.449..848H, 2016MNRAS.460.4188G, 2017ApJ...848...76Z}.
    The green and orange bands indicate the $68\%$ confidence contours of the evolution
    (eqn.~\ref{equation::result_evolution_bias_qso_galaxy})
    fit to the data point.
    Right: evolution for the three Mg biases, $b_{i}$, from the results
    of individual fit to the cross-correlations (first part of table~\ref{table::result_fit}).
    The orange, green and blue bands produce
    the contours from the fit to all orange, green and blue data points (eqn.~\ref{equation::result_bias_MG}).
    The results for the three combined fit to the
    cross-correlations are given in red ($z_{\mathrm{eff}} \in \{0.55,0.92,0.59\} $,
    two last parts of table~\ref{table::result_fit}).
    }
    \label{figure::evolution_bias_qso_galaxy__evolution_bias}
\end{figure*}

We model the evolution of transmission and halo bias by the following power-law:
\begin{equation}
    b_{i}(z) =
    b_{i}(z_{\mathrm{eff}})
    \left(
    \frac{1+z}{1+z_{\mathrm{eff}}}
    \right)^{\gamma_{i}}.
    \label{equation::evolution_bias}
\end{equation}
For quasars we adopt the measured values of the bias at different redshifts in
\citet{2005MNRAS.356..415C, 2013ApJ...778...98S, 2016JCAP...11..060L, 2017JCAP...07..017L}, and for galaxies in
\citet{2015MNRAS.449..848H, 2016MNRAS.460.4188G, 2017ApJ...848...76Z}.
All these results are presented in the left panel of
figure~\ref{figure::evolution_bias_qso_galaxy__evolution_bias}, after correction
for the different assumptions of fiducial
cosmology\footnote{\href{http://cosmocalc.icrar.org/}{http://cosmocalc.icrar.org/}}.
The two free parameters of equation~\ref{equation::evolution_bias}
for both tracers are determined through a fit of those measurements,
assuming they are independent:
\begin{equation}
    \begin{array}{lllllll}
        b_{\mathrm{quasar}}(z_{\mathrm{eff}} = 0.59) = 1.24 \pm 0.05,\\
        \gamma_{\mathrm{quasar}} = 1.44 \pm 0.08,\\
        Corr = -93\%,\\
        \\
        b_{\mathrm{galaxy}}(z_{\mathrm{eff}} = 0.59) = 2.07 \pm 0.02,\\
        \gamma_{\mathrm{galaxy}} = 1.33 \pm 0.15,\\
        Corr = -18\%.
    \end{array}
    \label{equation::result_evolution_bias_qso_galaxy}
\end{equation}
Although the galaxies used in this study
are more biased than the quasars at the effective redshift $z_{\mathrm{eff}} = 0.59$,
the redshift evolution for both tracers is compatible.
Because of the degeneracies between magnesium and object bias, we fix
the galaxy and quasar bias and their evolution
as given by equation~\ref{equation::result_evolution_bias_qso_galaxy}
at the effective redshift of each cross-correlation.
We leave free the bias of the three different magnesium transitions, and
assume that their redshift evolution is given by the same power-law index
as the galaxies and we fix this parameter: $\gamma_{\mathrm{Mg}} = 1.33$.
This approach corresponds to no evolution, $\gamma_{\tau_{\mathrm{Mg\,m}}} = 0$,
of their optical depth of
equation~\ref{equation::def_transition_bias}: $\tau_{m}(z) = \tau_{m}$.
This assumption has no consequences on the BAO measurement
(table~\ref{table::result_fit_systematics}), but affects the measurement
of the magnesium bias.
We revisit this point in section~\ref{subsection::Measure_of_the_Magmesium_bias}.

The RSD parameters for quasars and galaxies are given by the product:
\begin{equation}
    b_{i}(z_{\mathrm{eff}}) \beta_{i}(z_{\mathrm{eff}}) = f(z_{\mathrm{eff}}),
    \label{equation::relation_bias_beta_f}
\end{equation}
where $f$ represents the linear growth rate of structure. In \mbox{$\Lambda$CDM} cosmology
this quantity is approximated by: $f(z) = \Omega_{m}^{0.55}(z)$.
As with the bias of objects, we fix $f$ at the effective redshift of the different
cross-correlations.
The RSD parameter of Mg is highly correlated with the bias,
we therefore fix it to be equal to the RSD parameter of galaxies,
$\beta_{\mathrm{Mg}} = \beta_{\mathrm{galaxy}}$, for all three transitions.
We thus assume that MgII absorbers lie in galaxy halos.
Equation~\ref{equation::relation_bias_beta_f} also applies to the host
halos of the three magnesium transitions. Their optical
depth is given by:
\begin{equation}
    \tau_{m}(z) = - \frac{(b\beta)_{m} }{ f }(z).
    \label{equation::relation_tau_bias_beta_f}
\end{equation}

The quasi-linear power spectrum $P_{QL}$ of equation~\ref{equation::power_spectrum}
(eqn.~6.6 of Blomqvist2018)
is computed using CAMB \citep{2000ApJ...538..473L} and
depends on the BAO parameter: $\alpha_{\mathrm{iso}}$.
This parameter acts as an isotropic shift of the BAO wiggles along the
wavenumber $k$, corresponding to an isotropic shift of the BAO scale along
the direction $r$ of the correlation function \citep{2013JCAP...03..024K}.
We use a box prior of $\alpha_{\mathrm{iso}} \in [0.5,1.5]$.

The last two relevant parameters to our study are the overall shift of
the cross-correlation due to systematic errors in the
measurement of quasar and galaxy redshifts ($\Delta r_{\parallel}$: eqn.~6.19 of Blomqvist2018),
and the effect of statistical error in redshift measurement and non-linear
velocities ($\sigma_{v}$: eqn.~6.10 of Blomqvist2018).
Because some of our measurements offer weaker constraining power, for example when using the
Ly$\beta$ forest, we add a box prior on the redshift measurement parameter,
$\sigma_{v} \in [0, 10] \, h^{-1} \, \mathrm{Mpc}$, approximately corresponding to a maximum
error of $1000\,\mathrm{km\,s^{-1}}$.
This modification
affects only poorly measured correlations and has no effect on the measurement of BAO.
We remove this prior when performing the combined fit to all $30$ cross-correlations.

\vspace{10mm}
\subsection{Fit to the cross-correlations}
\label{section::Fit_to_the_cross_correlations}

The full model is composed of six free parameters.
Four parameters are the main focus of this study:
the cosmological BAO parameter, $\alpha_{\mathrm{iso}}$,
and the bias parameter of the metal transitions,
$b_{\mathrm{MgII(2796)}}$,
$b_{\mathrm{MgII(2804)}}$,
and
$b_{\mathrm{MgI(2853)}}$.
The two other nuisance parameters describe the redshift error distribution:
the systematic error, $\Delta r_{\parallel}$, and its width, $\sigma_{v}$.

All fits to the cross-correlation functions are done for a separation
$r \in \left[ 10,160 \right] \, h^{-1} \, \mathrm{Mpc}$ and for
$\mu \in [-1,1]$.
We fit each of the 30 different cross-correlations independently and list
the results in table~\ref{table::result_fit} at the effective redshift of each measurement.
The first part of the table presents the results for all 15 cross-correlations involving galaxies and for all
15 cross-correlations involving quasars.
The second part of this table presents the
combined fit using all $15$ galaxy cross-correlations to simultaneously
constrain the free parameters. We do the same for all quasar cross-correlations.
Finally, the third part of the table gives the combined fit
to all 30 different cross-correlations.

\begin{table*}
    \centering
    \scalebox{0.98}{
    \begin{tabular}{llllllllll}

$\mathrm{Correlation}$                    &
$z_{\mathrm{eff}}$                        &
$\alpha_{\mathrm{iso}}$                   &
$b_{\mathrm{MgII(2796)}}$ &
$b_{\mathrm{MgII(2804)}}$ &
$b_{\mathrm{MgI(2853)}}$  &  
$\chi^{2}_{\min{}}/DOF, probability$      \\ 

[tracer, forest]&
&
&
$\left[ 10^{-4} \right]$ &
$\left[ 10^{-4} \right]$ &
$\left[ 10^{-4} \right]$ &  
\\

    \noalign{\smallskip}
    \hline \hline
    \noalign{\smallskip}
$\mathrm{Galaxy,Ly\beta}$  & $0.44$ & $-$      & $\,\,\,-3.7 \pm 7.3$         & $\,\,\, -6.9 \pm 5.2$       & $\,\,\,\,\,\,\, 5.3 \pm 5.5$        & $2474.63 / (2504-6),  p = 0.63$ \\ 
$\mathrm{Galaxy,Ly\alpha}$ & $0.44$ & $-$      & $\,\,\,-2.1 \pm 3.5$         & $\,\,\, -6.3 \pm 3.2$       & $\,\,\,\,\,\,\, 2.6 \pm 2.0$        & $2604.66 / (2504-6),  p = 0.067$ \\ 
$\mathrm{Galaxy,SiIV}$     & $0.53$ & $-$      & $\,\,\,-3.6 \pm 1.9$         & $\,\,\, -6.6 \pm 1.9$       & $-1.08 \pm 0.68$            & $2503.98 / (2504-6),  p = 0.46$ \\ 
$\mathrm{Galaxy,CIV} $     & $0.55$ & $-$           & $\,\,\,-6.5 \pm 1.8$         & $\,\,\, -4.9 \pm 1.2$       & $\,\,\, -2.2 \pm 0.78$      & $2467.21 / (2504-6),  p = 0.67$ \\ 
$\mathrm{Galaxy,MgII(1)}$  & $0.54$ & $-$      & $\,\,\,-6.6 \pm 1.6$         & $\,\,\, -3.1 \pm 1.4$       & $-0.78 \pm 0.90$            & $2641.76 / (2504-6),  p = 0.022$ \\ 
$\mathrm{Galaxy,MgII(2)}$  & $0.55$ & $-$           & $\,\,\,-6.1 \pm 2.7$         & $\,\,\, -5.2 \pm 2.6$       & $\,\,\, -1.6 \pm 1.0$       & $2591.88 / (2504-6),  p = 0.093$ \\ 
$\mathrm{Galaxy,MgII(3)}$  & $0.55$ & $-$      & $\,\,\,-5.8 \pm 1.8$         & $\,\,\, -3.6 \pm 1.1$       & $-0.25 \pm 0.73$            & $2423.86 / (2504-6),  p = 0.85$ \\ 
$\mathrm{Galaxy,MgII(4)}$  & $0.55$ & $-$           & $\,\,\,-6.3 \pm 2.1$         & $\,\,\, -5.6 \pm 1.5$       & $-1.44 \pm 0.89$            & $2576.22 / (2504-6),  p = 0.13$ \\ 
$\mathrm{Galaxy,MgII(5)}$  & $0.55$ & $-$      & $\,\,\,-0.8 \pm 3.3$         & $-11.4 \pm 3.9$             & $\,\,\, -3.2 \pm 1.3$       & $2589.02 / (2504-6),  p = 0.10$ \\ 
$\mathrm{Galaxy,MgII(6)}$  & $0.55$ & $-$           & $\,\,\,-9.0 \pm 2.0$         & $\,\,\, -5.5 \pm 1.8$       & $-2.63 \pm 0.96$            & $2431.56 / (2504-6),  p = 0.83$ \\ 
$\mathrm{Galaxy,MgII(7)}$  & $0.55$ & $-$      & $\,\,\,-2.7 \pm 2.8$         & $\,\,\, -8.1 \pm 2.4$       & $-1.16 \pm 0.86$            & $2433.85 / (2504-6),  p = 0.82$ \\ 
$\mathrm{Galaxy,MgII(8)}$  & $0.55$ & $-$      & $\,\,\,\,\,\,\, 1.2 \pm 3.7$ & $-15.2 \pm 3.1$             & $\,\,\, -1.8 \pm 1.6$       & $2375.98 / (2504-6),  p = 0.96$ \\ 
$\mathrm{Galaxy,MgII(9)}$  & $0.57$ & $-$      & $\,\,\,-7.6 \pm 1.9$         & $\,\,\, -5.1 \pm 1.7$       & $-2.28 \pm 0.87$            & $2542.47 / (2504-6),  p = 0.26$ \\ 
$\mathrm{Galaxy,MgII(10)}$ & $0.57$ & $-$      & $\,\,\,\,\,\,\, 3.6 \pm 4.4$ & $-19.7 \pm 3.4$             & $\,\,\, -4.2 \pm 1.8$       & $2729.40 / (2504-6),  p = 7.2 \times 10^{-4}$ \\ 
$\mathrm{Galaxy,MgII(11)}$ & $0.58$ & $-$           & $\,\,\,-9.4 \pm 2.3$         & $\,\,\, -5.8 \pm 2.7$       & $-0.21 \pm 0.89$            & $2463.90 / (2504-6),  p = 0.68$ \\ 

\noalign{\smallskip}
\noalign{\smallskip}

$\mathrm{Quasar,Ly\beta}$  & $0.47$ & $-$      & $\,\,\,\, 45.0 \pm 52.0$    & $-78.0 \pm 44.0$            & $\,\,\,-9.0 \pm 36.0$      & $2476.67 / (2504-6),  p = 0.62$ \\ 
$\mathrm{Quasar,Ly\alpha}$ & $0.50$ & $-$      & $\,\,\, -7.0 \pm 13.0$      & $\,\,\,\,11.4 \pm 8.3$      & $\,\,\,-2.0 \pm 7.9$       & $2506.20 / (2504-6),  p = 0.45$ \\ 
$\mathrm{Quasar,SiIV}   $  & $0.68$ & $-$      & $\,\,\, -8.3 \pm 7.8$       & $-10.9 \pm 8.3$             & $\,\,\,-4.8 \pm 4.4$       & $2470.41 / (2504-6),  p = 0.65$ \\ 
$\mathrm{Quasar,CIV}    $  & $0.77$ & $-$      & $-11.3 \pm 4.9$             & $\,\,\,-4.5 \pm 4.0$        & $\,\,\,-2.7 \pm 2.9$       & $2593.62 / (2504-6),  p = 0.089$ \\ 
$\mathrm{Quasar,MgII(1)}$  & $0.85$ & $-$      & $\,\,\, -6.6 \pm 4.6$       & $-10.9 \pm 3.8$             & $\,\,\,-6.6 \pm 2.4$       & $2637.83 / (2504-6),  p = 0.025$ \\ 
$\mathrm{Quasar,MgII(2)}$  & $0.88$ & $-$      & $-20.4 \pm 6.0$             & $\,\,\,-1.8 \pm 6.5$        & $\,\,\,-5.4 \pm 3.5$       & $2616.77 / (2504-6),  p = 0.048$ \\ 
$\mathrm{Quasar,MgII(3)}$  & $0.92$ & $-$           & $\,\,\, -9.3 \pm 5.8$       & $-10.5 \pm 5.2$             & $\,\,\,\,\,\,\,0.4 \pm 2.7$        & $2540.53 / (2504-6),  p = 0.27$ \\ 
$\mathrm{Quasar,MgII(4)}$  & $0.94$ & $-$           & $\,\,\, -9.7 \pm 7.2$       & $-10.1 \pm 7.0$             & $\,\,\,-3.5 \pm 3.9$       & $2638.80 / (2504-6),  p = 0.025$ \\ 
$\mathrm{Quasar,MgII(5)}$  & $0.95$ & $-$      & $-26.8 \pm 6.4$             & $\,\,\,\,\,\,\,3.8 \pm 7.1$ & $\,\,\,\,\,\,\,0.8 \pm 3.8$        & $2397.27 / (2504-6),  p = 0.92$ \\ 
$\mathrm{Quasar,MgII(6)}$  & $0.96$ & $-$      & $-13.5 \pm 6.2$             & $\,\,\,-8.0 \pm 4.8$        & $\,\,\,-7.6 \pm 4.2$       & $2252.95 / (2504-6),  p = 1.00$ \\ 
$\mathrm{Quasar,MgII(7)}$  & $0.96$ & $-$      & $-15.7 \pm 7.2$             & $-14.9 \pm 7.2$             & $\,\,\,-6.7 \pm 3.1$       & $2413.65 / (2504-6),  p = 0.88$ \\ 
$\mathrm{Quasar,MgII(8)}$  & $0.97$ & $-$           & $\,\,\, -9.1 \pm 5.1$       & $-10.2 \pm 5.0$             & $\,\,\,-4.4 \pm 2.8$       & $2410.16 / (2504-6),  p = 0.89$ \\ 
$\mathrm{Quasar,MgII(9)}$  & $0.97$ & $-$      & $-23.0 \pm 6.3$             & $-11.7 \pm 6.2$             & $\,\,\,-1.1 \pm 2.8$       & $2478.24 / (2504-6),  p = 0.61$ \\ 
$\mathrm{Quasar,MgII(10)}$ & $0.96$ & $-$      & $-13.8 \pm 5.4$             & $-13.7 \pm 6.7$             & $\,\,\,-0.3 \pm 3.3$       & $2598.24 / (2504-6),  p = 0.079$ \\ 
$\mathrm{Quasar,MgII(11)}$ & $1.03$ & $-$           & $-26.0 \pm 14.0$            & $\,\,\,-5.7 \pm 8.6$        & $\,\,\,-3.4 \pm 3.6$       & $2549.83 / (2504-6),  p = 0.23$ \\ 

    \noalign{\smallskip}
    \hline
    \noalign{\smallskip}

$\mathrm{Galaxy,All}$ & $0.55$ & $0.982 \pm 0.049$    & $\,\,\, -6.2 \pm 0.55$      & $-4.99 \pm 0.50$     & $-1.08 \pm 0.22$     & $37916.30 / (37560-6),  p = 0.093$ \\ 
\noalign{\smallskip}
\noalign{\smallskip}
$\mathrm{QSO,All}    $ & $0.92$ & $1.018 \pm 0.052$    & $-14.3 \pm 1.8$      & $\,\,\,-7.9 \pm 1.5$       & $-2.77 \pm 0.85$     & $37655.48 / (37560-6),  p = 0.35$ \\ 

    \noalign{\smallskip}
    \hline
    \noalign{\smallskip}

$\mathrm{All,All}    $ & $0.59$ & $0.997 \pm 0.037$    & $-7.32 \pm 0.57$     & $-5.28 \pm 0.58$     & $-1.18 \pm 0.21$     & $75597.92 / (75120-6),  p = 0.11$ \\

    \noalign{\smallskip}

    \end{tabular}
    }
    \caption{
    Best fit parameters of the BAO and the three Mg biases (for visualization purposes
    the biases are multiplied by $10^{4}$).
    The first section lists the results for individual fit to
    each cross-correlations.
    The second section gives the results for the combined fit to all $15$ cross-correlations
    involving galaxies, then to all $15$ cross-correlations involving quasars.
    The third section presents the results for the combined fit to all $30$ cross-correlations.
    Since each individual fit doesn't constrain the BAO parameter, only the
    results for combined fits are shown.
    }
    \label{table::result_fit}
\end{table*}

For cross-correlations involving galaxies, the effective redshifts lie
in a relatively small range: from $z_{\mathrm{eff}}=0.44$ to $z_{\mathrm{eff}}=0.58$. For quasars, however,
the effective redshifts cover a larger range: from $z_{\mathrm{eff}}=0.47$ to $z_{\mathrm{eff}}=1.03$.
The combined fit to all cross-correlations involving galaxies has an
effective redshift $z_{\mathrm{eff}} = 0.55$, and for quasars it is $z_{\mathrm{eff}} = 0.92$.
If the bias evolves with redshift, we do not expect its best fit
value to agree between bins of different effective redshift.

In this table, the errors are given as the second derivative at the minimum,
evaluated at the extrapolated $\Delta \chi^{2} = 1$.
They do not exactly correspond to direct assessment of $\Delta \chi^{2} = 1$, nor to $68.27\%$ of trials.
The values of $b_{i}$ are of order $10^{-4}$; for clarity in this table
we multiply them by $10^{4}$.
The BAO parameter can not be measured significantly in each individual correlation,
we therefore present only the best fit results when combining the different measurement:
last three lines of table~\ref{table::result_fit}.

Among the $30$ individual cross-correlations, $27$ have probabilities $0.0228 < p < 0.977$, 
corresponding to two sigma significance, slightly fewer than the $29$ that would be expected
from this sample size. The galaxy-MgII(10) cross-correlation has an extremely low probability of $\chi^{2}$.
This aspect is explained by the estimation of the correlation-matrix of each
of the individual cross-correlation, not to the estimation of variance.
As explained in section~\ref{subsection::The_covariance_matrix},
because of numerical issues, we replaced the correlation matrix of
galaxy-MgII(8), quasar-MgII(8) and galaxy-MgII(10) by that of their neighboring
cross-correlation.
This action explains the low probability of $\chi^{2}$ for the galaxy-MgII(10) cross-correlation.
This result has little effect on the measurement of the best fit parameters and
errors.
We test this assumption in the last line of table~\ref{table::result_fit_systematics}.

We present in the right panel of 
figure~\ref{figure::corr_2d_all_quasar__all_galaxy__cutted} the stack
of all the best fit models, after running the combined fit to the
30~cross-correlations. The correlation
appears shifted towards positive values of $r_{\parallel}$.
This apparent feature is linked to the presence of \mbox{MgII(2804)} at
$r_{\parallel} \approx +9\,h^{-1} \, \mathrm{Mpc}$ of \mbox{MgII(2796)}
(table~\ref{table::metals_contribution}).
The doublet nature is the main source for the asymmetry between the
positive and the negative values
of $r_{\parallel}$.
At $(r_{\parallel},r_{\perp}) \approx (+68,0)\,h^{-1} \, \mathrm{Mpc}$, we observe the weaker
\mbox{MgI(2853)} correlation.

\begin{figure*}
    \centering
    \includegraphics[width=0.98\columnwidth]{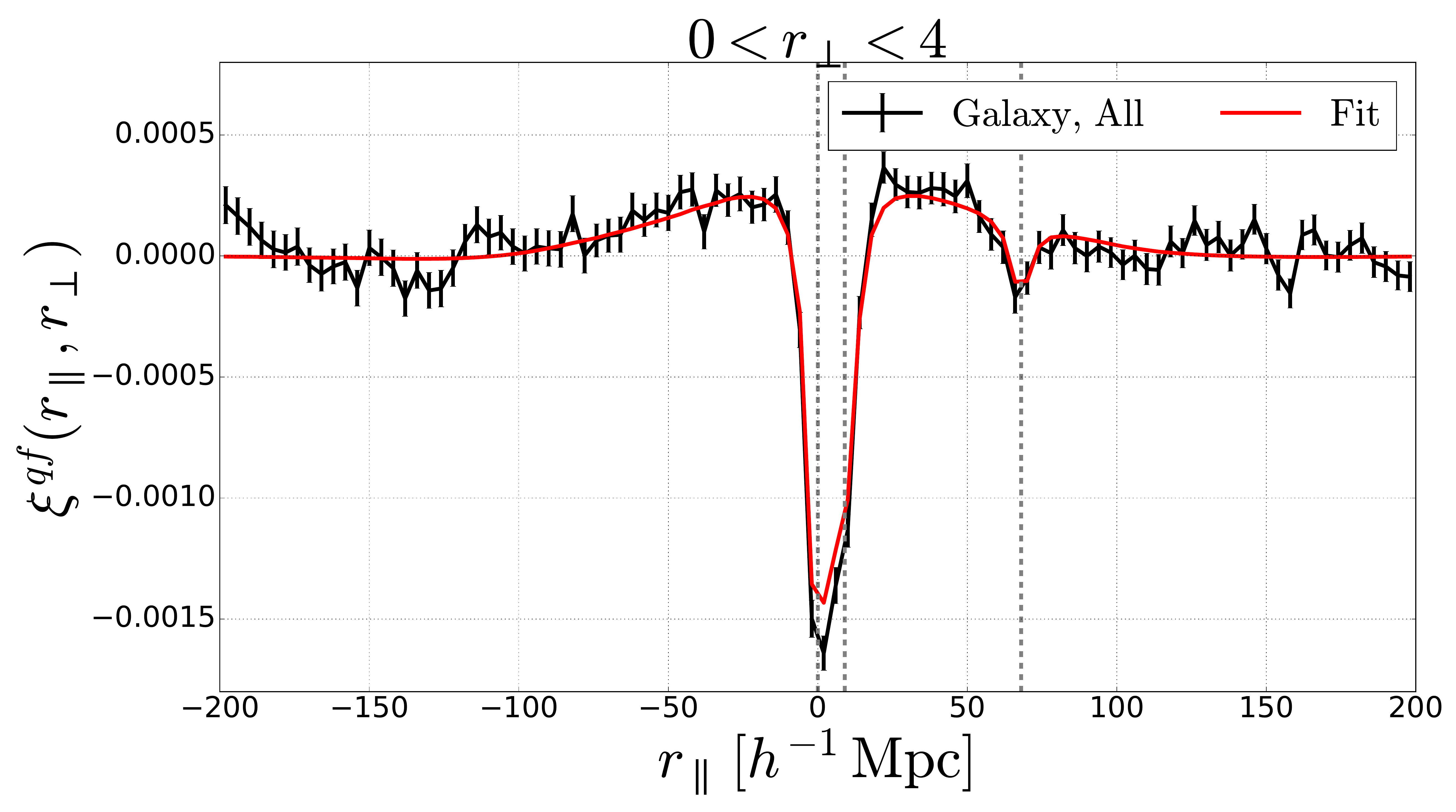}
    \includegraphics[width=0.98\columnwidth]{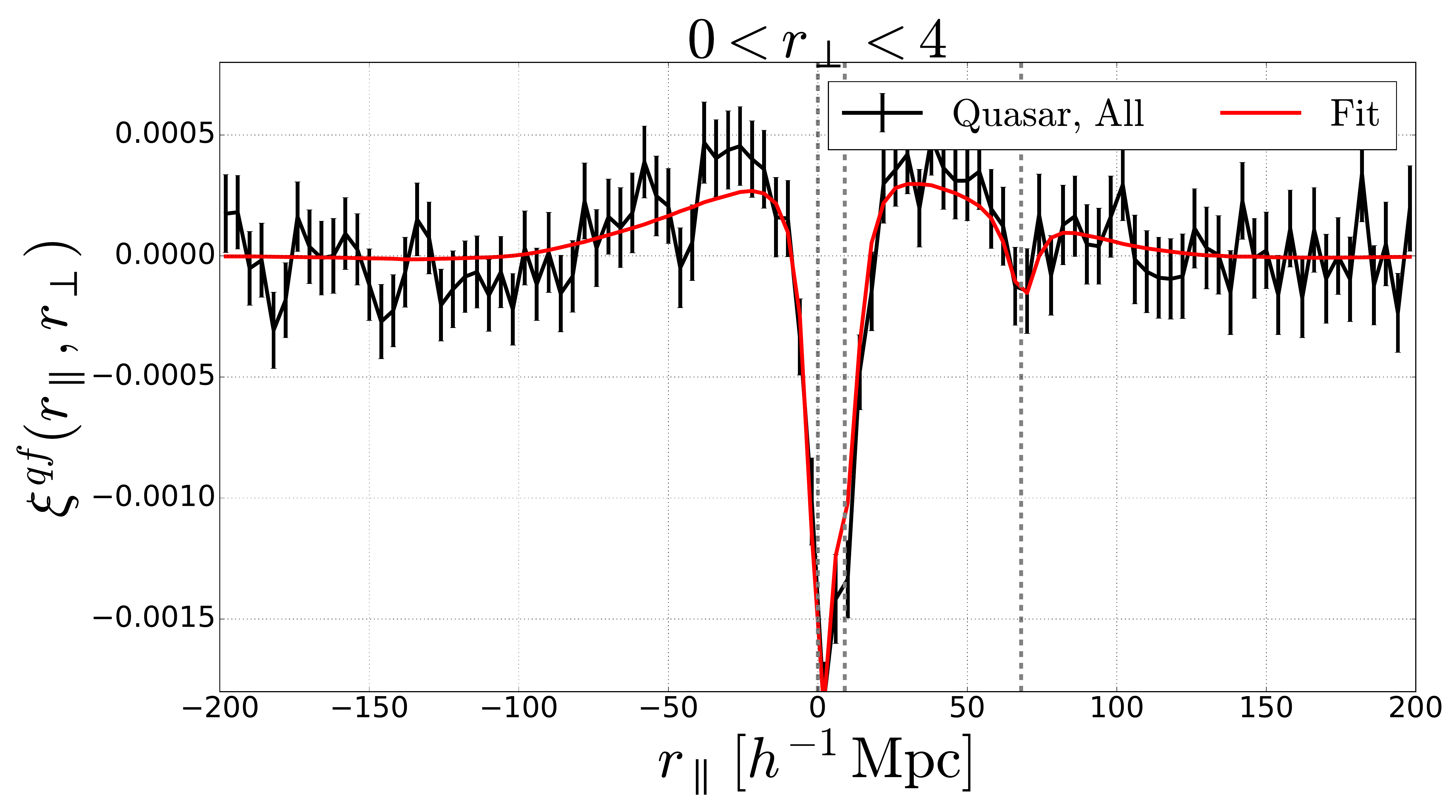}
    \includegraphics[width=0.98\columnwidth]{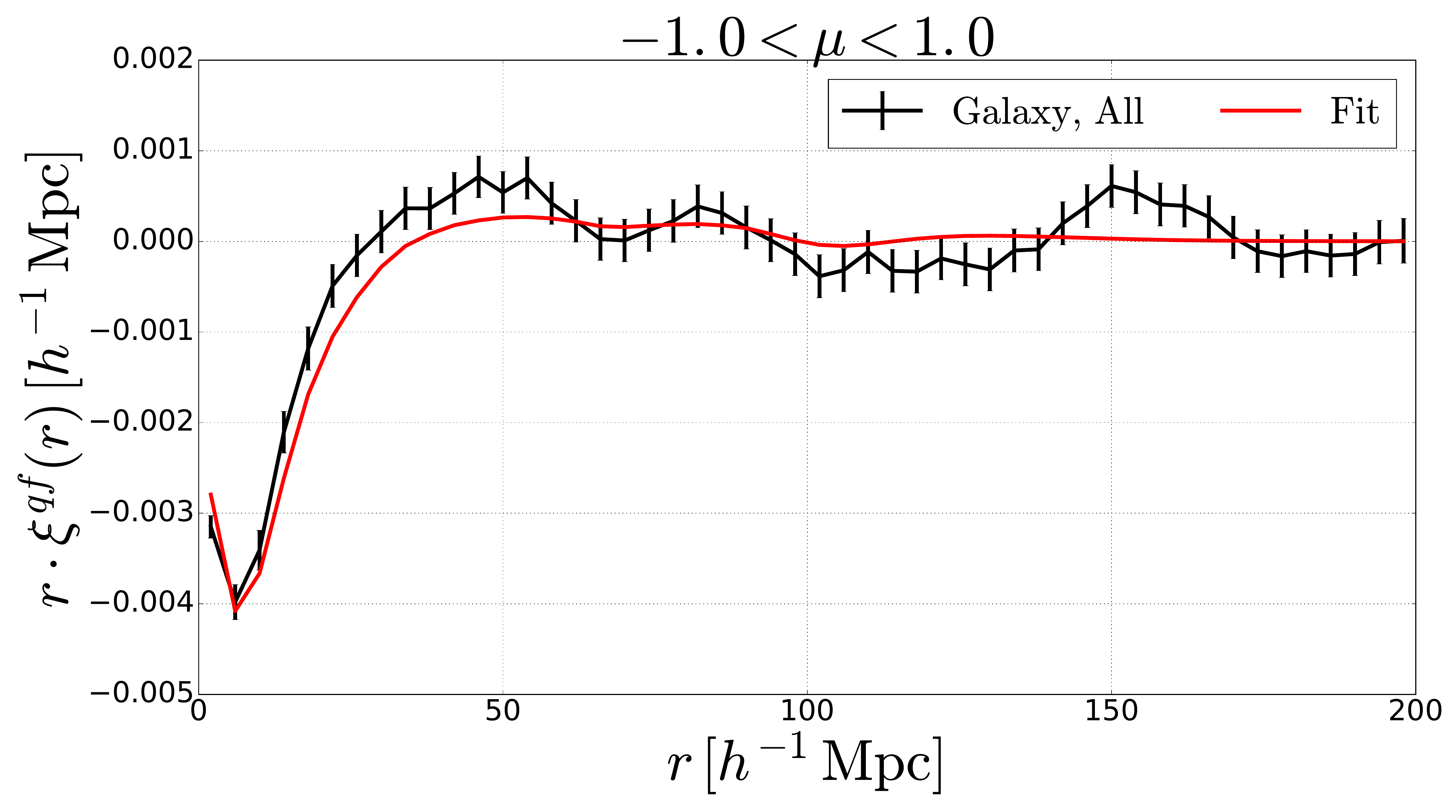}
    \includegraphics[width=0.98\columnwidth]{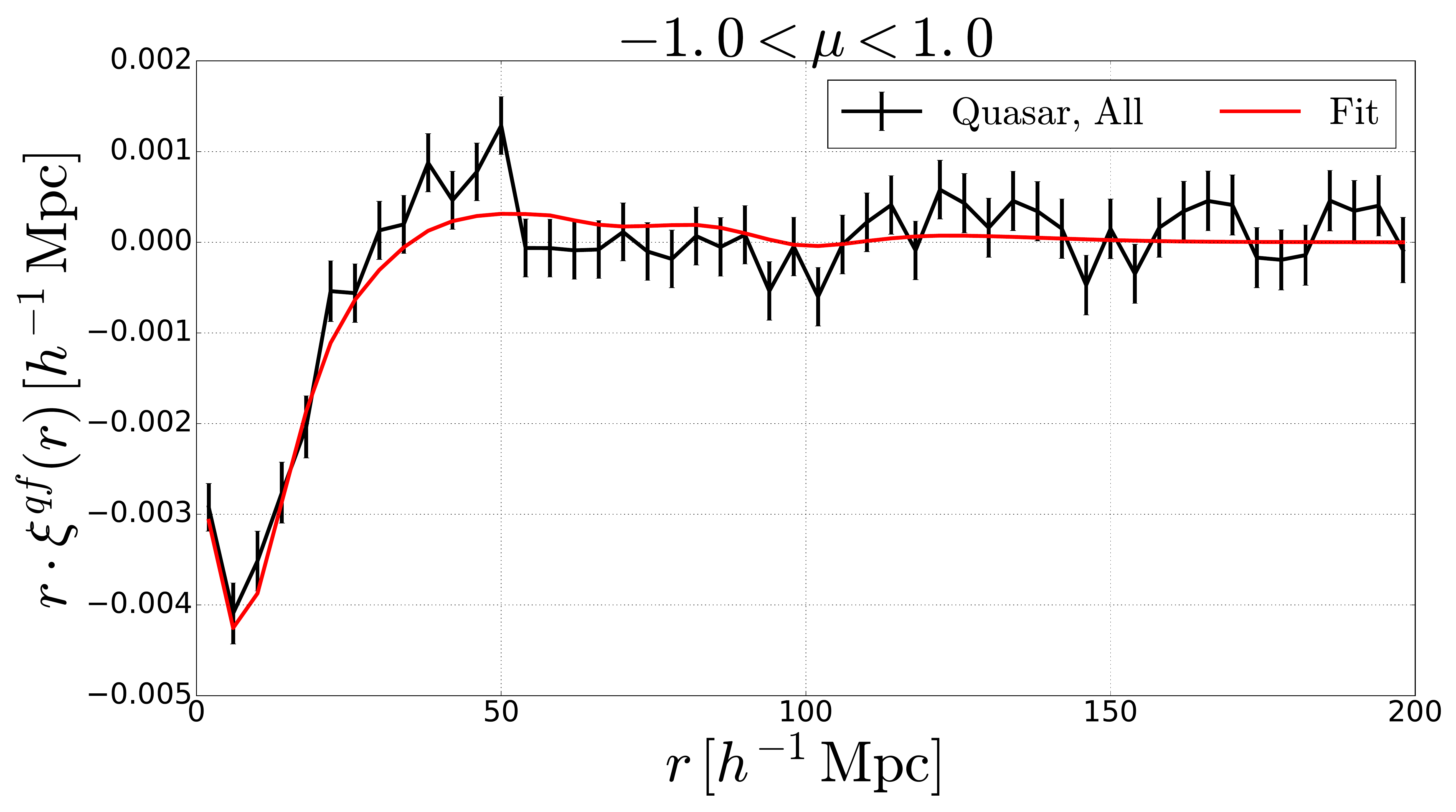}
    \includegraphics[width=0.98\columnwidth]{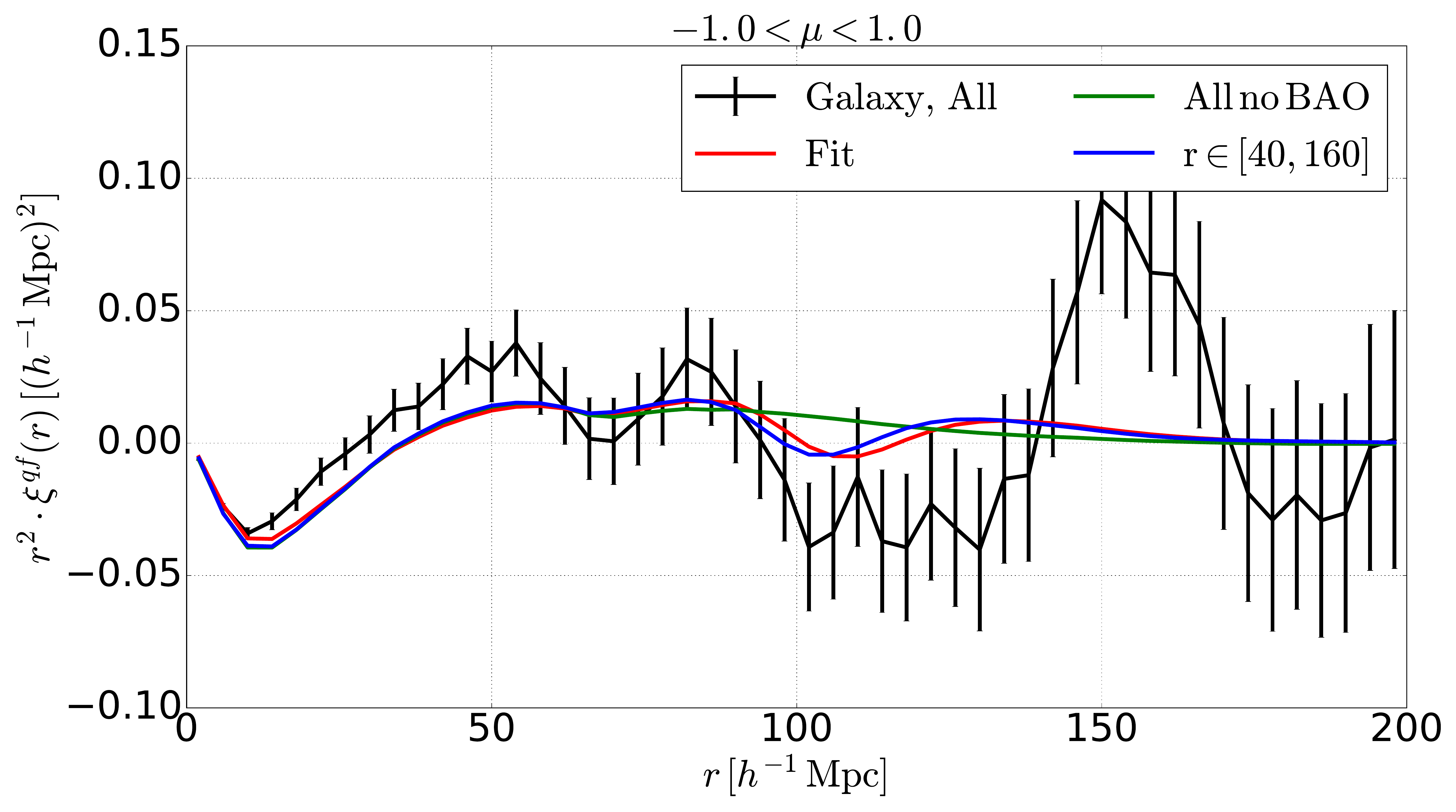}
    \includegraphics[width=0.98\columnwidth]{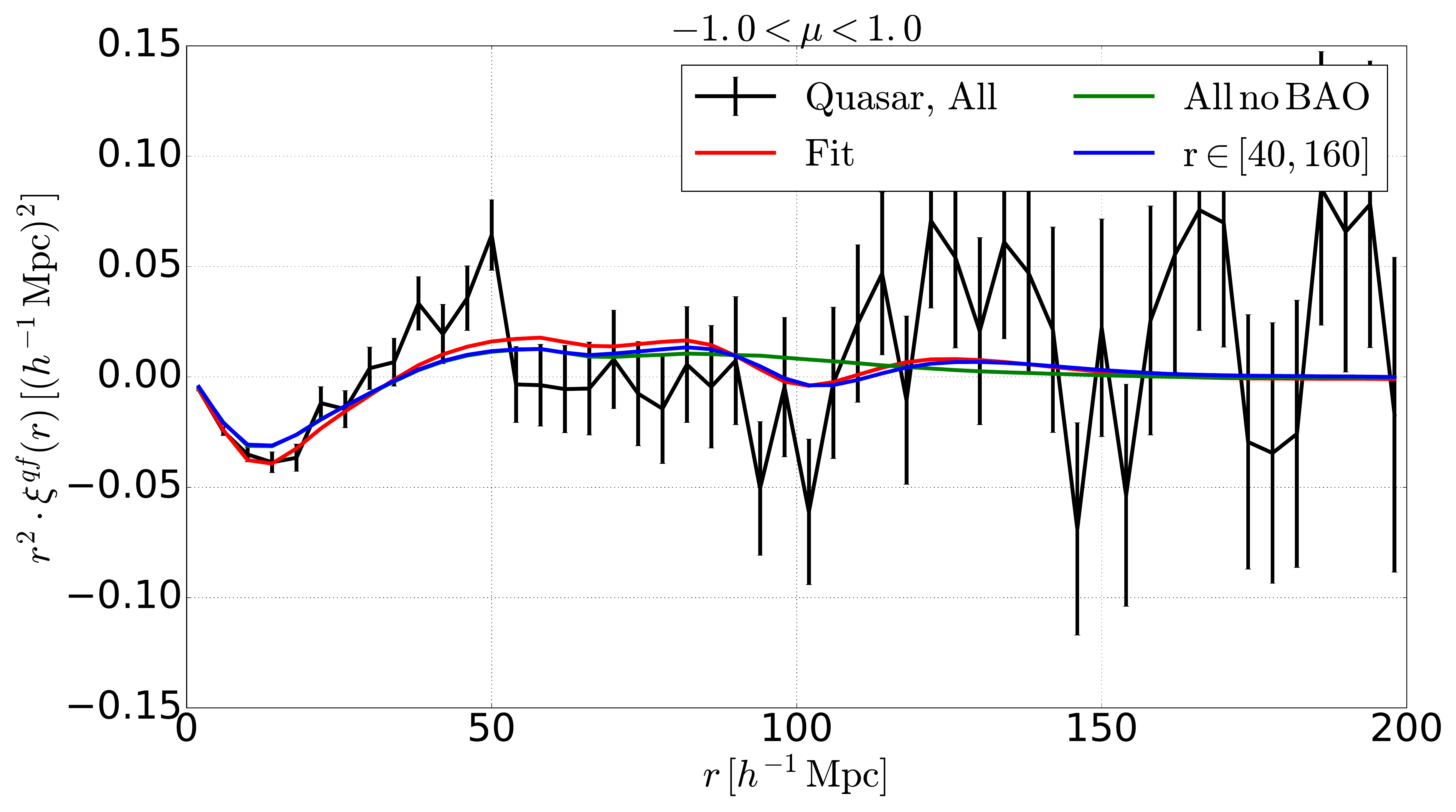}
    \caption{
    Comparison between the best fit and the data of the MgII object cross-correlations.
    Left (right) panels display the stacked best fit and data of all
    15 MgII galaxy (quasar) cross-correlations.
    The top two panels present the cross-correlations for pairs with
    $r_{\perp} \approx 0 \, h^{-1}\,\mathrm{Mpc}$ (small angular separation).
    The three different Mg correlation maxima are outlined by the gray
    dashed lines.
    The middle two panels show the spherically averaged correlation function,
    multiplied by the absolute separation $r$.
    The bottom panels show the same correlations, multiplied
    by the absolute separation $r^{2}$. The standard fit is shown in red,
    a fit without the BAO feature in green and a fit of bins in $[40,160]\,\hMpc$
    instead of $[10,160]\,\hMpc$ in blue.
    }
    \label{figure::slice_QSO_0__monopole_QSO_1}
\end{figure*}

Figure~\ref{figure::slice_QSO_0__monopole_QSO_1} presents a comparison
of the stacked data to the stacked best fit model, for galaxies on the left
and quasars on the right. The top panels present the correlation for pairs
with $r_{\perp} \approx 0 \, h^{-1} \, \mathrm{Mpc}$, i.e., pairs with a
small angular separation. These panels highlight the three metal-object
correlations. The two elements of the doublet, \mbox{MgII(2796)} and
\mbox{MgII(2804)}, are blended at
$r_{\parallel} \approx 0 \, h^{-1} \, \mathrm{Mpc}$.
At  $r_{\parallel} \approx +68\,h^{-1} \, \mathrm{Mpc}$, we observe the weaker
contribution of \mbox{MgI(2853)}.
These two top panels aim at presenting the fit of the three different correlations
of the three different Mg transitions; at our level of precision the BAO
feature can not be seen in such figures.
The middle panels give the spherically-averaged correlation for both
MgII-galaxy and MgII-quasar cross-correlations multiplied
by the absolute separation $r$.
Finally the bottom two panels show the same two correlations,
multiplied by the absolute separation $r^{2}$.
In these last two panels we give in red the mean standard fit,
in green the mean fit with no BAO feature and in blue a fit of bins in $[40,160]\,\hMpc$
instead of $[10,160]\,\hMpc$.
In the fit, the BAO peak can be observed at $r\sim100\,\hMpc$. In the data,
the BAO feature is only weakly statistically detected: $\Delta \chi^{2} = 7.25$
(sec.~\ref{subsection::Measure_of_the_Baryonic_Acoustic_Oscillations}).
Because of the important correlation between different bins of the cross-correlation
involving galaxies (figure~\ref{figure::covariance_matrix_01}) and the stack of the
fit and data, the fit does not go through the points at
$r<50\,\hMpc$, however the probability of $\chi^{2}$ is $11\%$.
In a similar way, the large scale fluctuations about the fit can be
explained by the large correlations of the bins of the correlation function.

\subsection{Measurement of the baryonic acoustic oscillations}
\label{subsection::Measure_of_the_Baryonic_Acoustic_Oscillations}

The measurements of BAO in each individual cross-correlation have an
average uncertainty of $18\%$.
However, the combined fit to all
30 correlations (last part of table~\ref{table::result_fit}) leads to
isotropic BAO constraints with better than $4\%$ precision.
The BAOs correlation with the other five parameters of our model
is small: less than $1\%$.

\begin{figure}
    \centering
    \includegraphics[width=0.98\columnwidth]{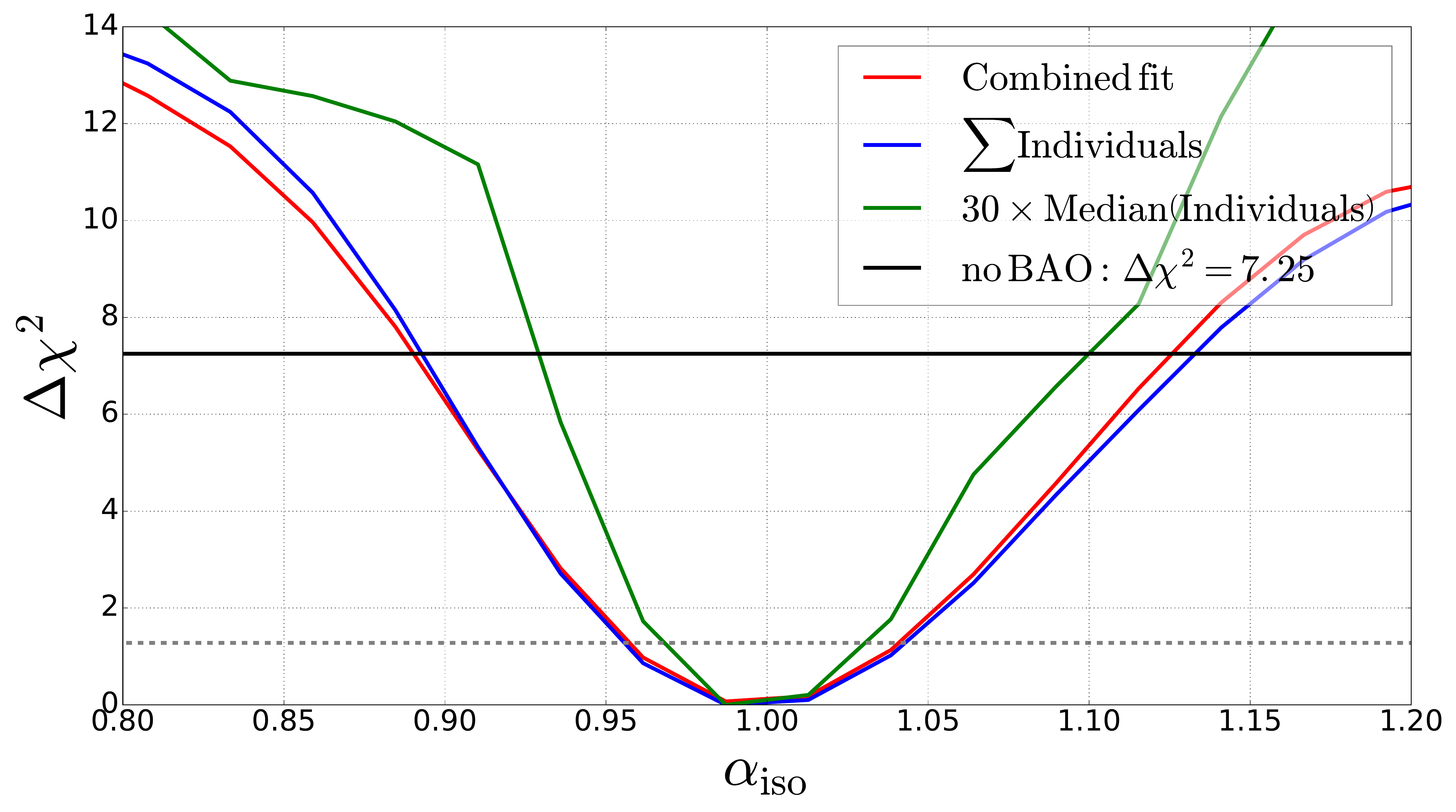}
    \caption{
    Result for the BAO parameter from the combined fit to all 30~cross-correlations
    (last line of table~\ref{table::result_fit}). 
    The $\Delta \chi^{2} = \chi^{2} - \chi_{\min}^{2}$ curve for the BAO parameter,
    $\alpha_{\mathrm{iso}}$,
    is in red for the combined fit,
    in blue for the sum of all individual fits,
    and in  green for the median of the $30$ individual fits multiplied by $30$.
    The gray dashed line indicates $\Delta \chi^{2} = 1.28$, corresponding to
    $\sigma = 68.27\%$ confidence levels.
    The black line is the $\Delta \chi^{2} = 7.25$ limit for a model without BAO
    using a combined fit to all $30$ cross-correlations.
    }
    \label{figure::chi2_scan_aiso__fastMC_alphaiso}
\end{figure}


Figure~\ref{figure::chi2_scan_aiso__fastMC_alphaiso} presents the
$\Delta \chi^{2} = \chi^{2} - \chi_{\min}^{2}$ curve around
the best fit $\alpha_{\mathrm{iso}}$.
The red curve gives the result for the combined fit, as shown in the bottom
of table~\ref{table::result_fit}.
The blue curve represents the result for the sum of all $30$ individual fits.
The green curve is the median of the $30$ individual fits multiplied by $30$.
The three uncertainties yield similar best fit values and errors, thus
providing evidence that the individual fits are robust even in the regime
of low signal-over-noise ratio.
The difference between the median and the combined fit is explained
by the large differences in statistics between the $30$~individual fits.

In table~\ref{table::result_fit_systematics} of appendix~\ref{section::Sytematic_tests},
we present different systematic tests on
the best fit results for $\alpha_{\mathrm{iso}}$ when changing the models
or the fitting range for the combined fit to the 30 cross-correlations.
No significant changes in the best fit value are detected.
We do find one change in the measurement precision by a factor
$1.7$ when the amplitude of the BAO peak is introduced as a free parameter.
The data are best described with a BAO peak of amplitude $A_{\mathrm{peak}} = 2.92 \pm 0.82$.
The fit using the peak as a free parameter results in
$\Delta \chi^{2} = 5.5$ from $A_{\mathrm{peak}} = 1$, corresponding to a less than $3 \sigma$ detection.
This enhancement of the BAO peak amplitude could be statistical, linked to
spurious signal, or the result of suppressed broadband shape in the measured correlation function.
We take the conservative option and keep this parameter fixed to its fiducial
value of $A_{\mathrm{peak}} = 1$.

To determine if one cross-correlation is driving the results of $\alpha_{\mathrm{iso}}$
and $A_{\mathrm{peak}}$, we compute the combined fit $30$ times removing
one of the correlations each time.
We perform a similar jackknife, removing also two cross-correlations involving
the same forest, producing $15$ different combined fits.
The BAO scale and peak size best fit values and errors are compatible with the
statistical precision.

The fact that no significant change of the BAO best fit is measured
in table~\ref{table::result_fit_systematics}, or in the jackknife,
allows us to assess the different
points of the introduction (sec.~\ref{section::Introduction}).
We discuss these points and their consequences for the Ly$\alpha$ analyses in the
conclusion (sec.~\ref{section::Summary_and_conclusions}).

The table in the appendix gives, in the last line of the second section,
the $\chi^{2}$ for a model with no BAO ($A_{\mathrm{peak}}=0$).
When this result is compared to our standard model ($A_{\mathrm{peak}}=1$),
it yields $\Delta \chi^{2} = 7.25$, shown as a black line in
figure~\ref{figure::chi2_scan_aiso__fastMC_alphaiso}.
This low significance of the BAO peak is consistent with the lack of an
evident BAO peak in figure~\ref{figure::slice_QSO_0__monopole_QSO_1}.
Such a fit is presented in figure~\ref{figure::slice_QSO_0__monopole_QSO_1}
by the green line in both bottom panels.
A similar significance of the BAO peak is obtained when fitting bins in $[40,160]\,\hMpc$,
see last section of table~\ref{table::result_fit_systematics}. The significance is then:
$\Delta \chi^{2} = 71171.24-71165.15 = 6.09$.

As described in dMdB2017, the fit of the BAO parameter is not linear.
The link between $\Delta \chi^{2}$ and $\sigma = 68.27\%$
must therefore be determined empirically. We determine the relation between BAO measurement
precision and the $\chi^{2}$ surface by using $100$ fast \mbox{Monte-Carlo}
(fastMC)
realizations of our measurements according to our best fit model and
covariance matrix. These $100$ fastMC realizations are fit leaving all six parameters free
and fixing $\alpha_{\mathrm{iso}} = 1$; this selections allows
one to efficiently create $100$ realizations
of $\Delta \chi^{2} = \chi^{2}_{\alpha_\mathrm{iso} = 1} - \chi^{2}_{\mathrm{all\,free}}$.
We find that $\Delta \chi^{2} = 1.28$ does indeed represent
$\sigma = 68.27\%$ of trials.

The final BAO measurement is generated by the combined fit to all $30$ correlations.
After the estimation of the relation between $\Delta \chi^{2}$ and confidence levels,
the measurement of the spherically-averaged BAO parameter is:
\begin{equation}
    \alpha_{\mathrm{iso}} = 0.997 \pm 0.047,
\end{equation}
where the error represents the $68.27\%$ confidence level.
This result is compatible with the cosmology of \citet{2016A&A...594A..13P}.

For comparison, \citet{2017MNRAS.470.2617A} measured
$\alpha_{\mathrm{iso}}$ with a $1\%$ error in each of the $z=0.32$ and $z=0.57$ bins
(their table~9) using the auto-correlation of galaxies from
\citet{2016MNRAS.455.1553R}.
Ours is the first measurement made at $z<2$ using MgII as a transmission
field to measure BAO parameters.

In their study of the CIV absorption in the Ly$\alpha$, SiIV and CIV forests,
Blomqvist2018 produced a similar measurement of the BAO parameter at $z=2$:
the significance of their BAO peak is given by $\Delta \chi^{2} = 3.22$ and
their measurement of $\alpha_{\mathrm{iso}}$ is at the $7\%$ level.
Only the studies of the Ly$\alpha$ absorption in the Ly$\alpha$ forest from
dMdB2017 and \citet{2017A&A...603A..12B} have a significant measurement
of the BAO peak at $z>2$, larger than that presented here,
respectively $\Delta \chi^{2} = 14$ and $\Delta \chi^{2} = 28$.
Although the CIV and MgII absorption fields are promising avenues for new BAO measurements,
they are not yet able to provide the same precision on the BAO distance scale
as the galaxy tracers or Ly$\alpha$ flux-transmission.
MgII and CIV do, however, probe the range $1<z<2$ which is
currently shot-noise limited in the two-year eBOSS quasar sample;
\citet{2018MNRAS.473.4773A} measures
$\alpha_{\mathrm{iso}} = 0.993 \pm 0.038$ at $z_{\mathrm{eff}} = 1.52$.

\subsection{Measurement of magnesium bias}
\label{subsection::Measure_of_the_Magmesium_bias}

For each of our $30$ individual cross-correlations, we have a measurement of the
bias, $b_{m}$, of the absorption field of MgII(2796), MgII(2804)
and MgI(2853) at their effective redshifts.
The results are given
in the first part of table~\ref{table::result_fit} and are presented in the right panel
of figure~\ref{figure::evolution_bias_qso_galaxy__evolution_bias}.
This panel gives in blue, green and orange the measurement of MgII(2796), MgII(2804)
and MgI(2853), respectively. The red dots are the measurements presented in
the last three lines of table~\ref{table::result_fit}, i.e., the
three different combined fits. At a redshift of $z=0.59$, the three red
points are for the combined fit to all of the $30$ cross-correlations.

These three bias measurements are correlated with one another and with
$\Delta r_{\parallel}$ and $\sigma_{v}$, but are only marginally correlated with
$\alpha_{\mathrm{iso}}$.
For the combined fit to all $30$ cross-correlations (last line of table~\ref{table::result_fit}),
$b_{\mathrm{MgII(2796)}}$ is correlated at the level of $-77\%$ with $b_{\mathrm{MgII(2804)}}$ and $\Delta r_{\parallel}$,
at the level of $46\%$ with $\sigma_{v}$,
and at the level of $-17\%$ with $b_{\mathrm{MgI(2853)}}$.
$b_{\mathrm{MgII(2804)}}$ is correlated at the level of $-89\%$ with $\Delta r_{\parallel}$,
at the level of $-64\%$ with $\sigma_{v}$,
and at the level of $25\%$ with $b_{\mathrm{MgI(2853)}}$.
$b_{\mathrm{MgI(2853)}}$ is correlated at the level of $-25\%$ with $\Delta r_{\parallel}$,
and at the level of $-26\%$ with $\sigma_{v}$.

We test for possible systematic errors in our measurement of the bias
of the three transitions and present the results in table~\ref{table::result_fit_systematics}
of appendix~\ref{section::Sytematic_tests}. The second section of the table
gives changes in the modeling of the BAO peak. Since the peak is decoupled
from the overall correlation function, we observe no changes in the Mg
bias measurements under varying assumptions of BAO.
In the third section, we modify the model of the fit to the cross-correlation,
and in the last section we modify the fitting range.
No significant changes in the best fit values of the bias parameters are
observed with the exception of three cases.

In the last two lines of the third section, we change the assumption on the
number and type of transitions observed: either we assume that MgII(2796) is a
singlet and that MgI(2853) is not present ($(b_{\mathrm{MgII(2804)}},b_{\mathrm{MgI(2853)}})=(0,0)$) or we
assume that MgII(2804) is a singlet ($b_{\mathrm{MgII(2796)}}=0$).
In the first line of the last section of the table we fit only bins
with $r_{\parallel}<0$.
The effect of these three changes is that the peak produced by the
MgII doublet at $r \approx 0 \, h^{-1} \, \mathrm{Mpc}$
is modeled as an MgII singlet - object cross-correlation.
The resulting effective MgII singlet transition has a bias equal
to the sum of the two components in the MgII doublet.
This expected value for the effective MgII singlet transition is
recovered in the three cases.
Thus, these three cases where we observe a significant difference with
the bias values of our standard fit are expected, and all yield a
bias compatible with $b_{\mathrm{MgII \, singlet}} = -12.6 \times 10^{-4}$.

Modeling a doublet as an effective singlet is done in Blomqvist2018
and in \citet{2018MNRAS.480..610G} for the CIV doublet correlation with
the quasar distribution.
This choice of analysis gives a measurement of an
effective bias of the transition that is the sum of the bias of the
two members of the doublet.
This approach was motivated by two aspects:
the CIV doublet has a smaller separation than the MgII doublet: $2.6$~\AA{} versus $7.2$~\AA{},
and this simplification has no effect on the BAO scale at this
level of precision (table~\ref{table::result_fit_systematics}).
This simplification of the MgII transition doublet into an effective singlet
has other consequences beyond the bias values.
It produces a model that describes the data with less significance.
In our study, the standard combined fit has $\chi^{2} = 75597.92$, while the fit
with only MgII(2804) and MgI(2853) ($b_{\mathrm{MgII(2796)}}=0$) has
$\chi^{2} = 75649.21$. The difference is $\Delta \chi^{2} = 51$,
corresponding to more than 5~$\sigma$ significance, for $1$ degree of freedom difference.
The effect occurs at small scales and does not bias estimates of the BAO scale.

The other two consequences of this assumption are not given in table~\ref{table::result_fit_systematics}.
First, modeling the doublet as a single line increases the value of $\sigma_{v}$,
the parameter representing the statistical error on the redshift of the
quasar or galaxies.
In our standard fit we measure $\sigma_{v} = 0.2 \pm 2.0 \, h^{-1} \, \mathrm{Mpc}$;
when modeling with a single line, $\sigma_{v} = 5.26 \pm 0.71 \, h^{-1} \, \mathrm{Mpc}$.
In a similar way, the parameter representing the systematic shift of the cross-correlations
due to biased redshift estimates is affected.
In our standard fit, we measure $\Delta r_{\parallel} = -0.06 \pm 0.43 \, h^{-1} \, \mathrm{Mpc}$;
when modeling with an effective line, $\Delta r_{\parallel} = 4.21 \pm 0.23 \, h^{-1} \, \mathrm{Mpc}$.
All of these aspects demonstrate the importance of modeling the transition properly as a
doublet.

We can not identify any major systematic errors in our measurement of the bias for the
three Mg transitions.
Contrary to the BAO parameter, $\alpha_{\mathrm{iso}}$, the relation between
$\Delta \chi^{2} = (1,4)$ and $(68.27\%,95.45\%)$ of trials is linear,
and requires no correction for the statistical uncertainty. 
Using all measurements at each effective
redshift from the first section of table~\ref{table::result_fit}, and taking
into account their correlation matrix, we fit the three bias values at
$z_{\mathrm{eff}}=0.59$ and a common redshift evolution parameter:
\begin{equation}
    \begin{array}{llll}
        b_{\mathrm{MgII(2796)}} (z = 0.59) = (-6.82 \pm 0.54) \, \times 10^{-4}, \\
        b_{\mathrm{MgII(2804)}} (z = 0.59) = (-5.55 \pm 0.46) \, \times 10^{-4}, \\
        b_{\mathrm{MgI(2853)}}  (z = 0.59) = (-1.48 \pm 0.24) \, \times 10^{-4}, \\
        \gamma_{\mathrm{Mg}} = 3.36 \pm 0.46. \\
    \end{array}
    \label{equation::result_bias_MG}
\end{equation}
The evolution parameter, $\gamma_{\mathrm{Mg}}$, defines the evolution
of each bias, $b_{m}$, as given in equation~\ref{equation::evolution_bias}.
A model with a different
evolution for each bias does not improve significantly the fit.
The three resulting biases are consistent with the values found
when performing a combined fit to all $30$ cross-correlations
(last line of table~\ref{table::result_fit}).
They also are compatible to fitting all $30$ cross-correlations, leaving
free $\gamma_{\mathrm{Mg}}$ (table~\ref{table::result_fit_systematics}).
From the third line of the third section of
table~\ref{table::result_fit_systematics}, we see that our baseline
assumption of $\gamma_{\mathrm{Mg}} = 1.33$ is disfavored at the level
of $\Delta \chi^{2} = 27$. This result suggests that $\gamma_{\tau_{\mathrm{Mg}}} \neq 0$,
i.e., the optical depth of magnesium from the IGM evolves with redshift.
This evolution has no consequences on the BAO best fit value.
The contribution of the error on the galaxy and quasar biases and redshift
evolution parameter from
equation~\ref{equation::result_evolution_bias_qso_galaxy} is negligible
on the result of equation~\ref{equation::result_bias_MG}.

The results of equation~\ref{equation::result_bias_MG} have correlations:
$b_{\mathrm{MgI(2853)}}$ is uncorrelated, but
$b_{\mathrm{MgII(2796)}}$ is $-19\%$ correlated with $b_{\mathrm{MgII(2804)}}$,
and $22\%$ with $\gamma_{\mathrm{Mg}}$;
$b_{\mathrm{MgII(2804)}}$ is $22\%$ correlated with
$\gamma_{\mathrm{Mg}}$. We present in the right panel of
figure~\ref{figure::evolution_bias_qso_galaxy__evolution_bias} the
$1\,\sigma$ band in blue, green and orange for the bias value and
evolution with redshift.

From equation~\ref{equation::relation_tau_bias_beta_f}, we can convert each
measurement of magnesium transmission bias to a measurement of
magnesium optical depth, and obtain the overall redshift evolution:
\begin{equation}
    \begin{array}{llll}
        \tau_{\mathrm{MgII(2796)}} (z = 0.59) = (3.37 \pm 0.25) \, \times 10^{-4}, \\
        \tau_{\mathrm{MgII(2804)}} (z = 0.59) = (2.64 \pm 0.21) \, \times 10^{-4}, \\
        \tau_{\mathrm{MgI(2853)}}  (z = 0.59) = (0.67 \pm 0.11) \, \times 10^{-4}, \\
        \gamma_{\tau_{\mathrm{Mg}}} =  2.07 \pm 0.31.\\
    \end{array}
    \label{equation::result_tau_MG}
\end{equation}


%
%
\section{Summary and conclusions}
\label{section::Summary_and_conclusions}

We measured the cross-correlation between the distribution
of quasars and galaxies with the absorption from \mbox{magnesium-II} in quasar spectra.
The measurement was performed using all available data from
SDSS-I through SDSS-IV, mostly from the BOSS and eBOSS programs.
It is the first time that this MgII-object cross-correlation has been
investigated on scales sufficiently large to measure the BAO feature.
We detect the correlation at high significance.

Our measurement yields a $4.7\%$ precision estimate of the isotropic BAO parameter
$\alpha_{\mathrm{iso}}$ at an effective redshift of $z_{\mathrm{eff}} = 0.59$.
At a similar redshift, the auto-correlation of galaxies from BOSS \citep{2017MNRAS.470.2617A}
constrains the same parameter with a $1\%$ precision in
two redshift bins.

The three magnesium bias parameters are:
$b_{\mathrm{MgII(2796)}} (z = 0.59) = (-6.82 \pm 0.54) \, \times 10^{-4}$,
$b_{\mathrm{MgII(2804)}} (z = 0.59) = (-5.55 \pm 0.46) \, \times 10^{-4}$,
and
$b_{\mathrm{MgI(2853)}}  (z = 0.59) = (-1.48 \pm 0.24) \, \times 10^{-4}$.
Their redshift evolution is characterized by the power-law index:
$\gamma_{\mathrm{Mg}} = 3.36 \pm 0.46$.

This analysis uses the same Python package, \texttt{picca}, used in Ly$\alpha$
forest BAO measurement in BOSS, eBOSS and in the upcoming
Dark Energy Spectroscopic Instrument \citep[DESI:][]{2016arXiv161100036D}.
This choice allows tests of the Ly$\alpha$ analyses methodology in the low signal, large
amount of data regime. The excellent agreement between the best fit model and
the data of the MgII - object cross-correlation demonstrates that \texttt{picca} enables fits of
complex correlations. For example, the main absorption lines,
MgII(2796) and MgII(2804), comprise a doublet,
and the other absorption line, MgI(2853), is relatively strong compared
to the strongest one, MgII(2796).
The fact that the three signatures appear to be properly modeled
demonstrates the robustness of the Ly$\alpha$ analyses
and its implementation in \texttt{picca}.

This study further demonstrates the robustness of the Ly$\alpha$ analyses,
as it was reported in \citet{2017A&A...603A..12B} and \citet{2017A&A...608A.130D}.
We further address three potential sources of systematic errors:
\begin{itemize}

    \item We define $15$ different forests and treat them independently
    to measure the absorption from
    MgII. We find no significant systematic errors on the BAO scale or
    on other model parameters, indications
    that the quasar unabsorbed continuum and its variations
    are correctly modeled.

    \item The model for the MgII doublet and MgI
    absorption results in a $\chi^{2}$ with probabilities that indicate
    that the model correctly describes the data.
    The three magnesium transitions are well modeled,
    as shown, e.g., in the two top panels of
    figure~\ref{figure::slice_QSO_0__monopole_QSO_1}.
    The tests also suggest that the effect of other metals are correctly modeled.
    
    \item The two members of the MgII doublet transition at $z=0.59$ each have a bias
    $\sim200$ times smaller than that of Ly$\alpha$ at $z=2.4$.
    This behavior makes our study more susceptible to systematic errors in flux
    calibration or sky residuals; however, we find no evidence for such
    errors.
    
\end{itemize}

This study also allows one to independently model the effect of the
auto-correlation of MgII embedded in the measured Ly$\alpha$ auto-correlation,
as it is done for the CIV auto-correlation embedded in the measured
Ly$\alpha$ auto-correlation of \citet{2017A&A...603A..12B}.

This study using MgII, and other analyses using CIV \citep[e.g.,][]{2018JCAP...05..029B},
open a new window toward measuring the
BAO scale at similar redshifts.
The completed eBOSS and DESI surveys will provide multiple low-redshift
quasars and galaxies necessary to improve
the precision on the BAO parameter from this approach.

%
%

\begin{acknowledgements}

We thank Pasquier Noterdaeme for providing the DLA catlog on eBOSS DR14
quasars.

The work of H\'elion du Mas des Bourboux, Kyle Dawson, and Vikrant Kamble was supported in
part by U.S. Department of Energy, Office of Science,
Office of High Energy Physics, under Award Number DESC0009959.

Funding for the Sloan Digital Sky Survey IV has been provided by the Alfred P. Sloan Foundation, the U.S. Department of Energy Office of Science, and the Participating Institutions. SDSS acknowledges support and resources from the Center for High-Performance Computing at the University of Utah. The SDSS web site is www.sdss.org.

SDSS is managed by the Astrophysical Research Consortium for the Participating Institutions of the SDSS Collaboration including the Brazilian Participation Group, the Carnegie Institution for Science, Carnegie Mellon University, the Chilean Participation Group, the French Participation Group, Harvard-Smithsonian Center for Astrophysics, Instituto de Astrofísica de Canarias, The Johns Hopkins University, Kavli Institute for the Physics and Mathematics of the Universe (IPMU) / University of Tokyo, the Korean Participation Group, Lawrence Berkeley National Laboratory, Leibniz Institut für Astrophysik Potsdam (AIP), Max-Planck-Institut für Astronomie (MPIA Heidelberg), Max-Planck-Institut für Astrophysik (MPA Garching), Max-Planck-Institut für Extraterrestrische Physik (MPE), National Astronomical Observatories of China, New Mexico State University, New York University, University of Notre Dame, Observatório Nacional / MCTI, The Ohio State University, Pennsylvania State University, Shanghai Astronomical Observatory, United Kingdom Participation Group, Universidad Nacional Autónoma de México, University of Arizona, University of Colorado Boulder, University of Oxford, University of Portsmouth, University of Utah, University of Virginia, University of Washington, University of Wisconsin, Vanderbilt University, and Yale University.

\end{acknowledgements}

%
%
\bibliographystyle{mnras}
\bibliography{mgIIDR14}

%
%
\appendix

\section{Systematic tests on BAO and magnesium bias}
\label{section::Sytematic_tests}

This appendix presents the set of tests on the combined fit to all $30$
different cross-correlations (last row of table~\ref{table::result_fit}).
We assess the best fit values and errors on BAO and magnesium biases 
under different assumptions in the model.
The results of all these tests are shown in table~\ref{table::result_fit_systematics}.
The impacts of these tests on our analysis are discussed in
section~\ref{subsection::Measure_of_the_Baryonic_Acoustic_Oscillations}
and~\ref{subsection::Measure_of_the_Magmesium_bias}.
The results demonstrate that our measurement is robust against different changes in the
analysis.

In this table, each row lists the best fit values and errors of
$\alpha_{\mathrm{iso}}$, 
$b_{\mathrm{MgII(2796)}}$,
$b_{\mathrm{MgII(2804)}}$,
$b_{\mathrm{MgI(2853)}}$ for one of the tests.
When a test has extra parameters, we present the best fit values for those parameters in
the first column.

The first section of this table (``Std'') recalls the best fit results in the model
chosen in this analysis.
This entry is a duplicate of the last row of table~\ref{table::result_fit}.
The other two parameters have best fit values of:
$\Delta r_{\parallel} = -0.06 \pm 0.43 \, \hMpc{}$ and
$\sigma_{v} = 0.2 \pm 2.0 \, \hMpc{}$.

The second section of this table presents the results of the tests on the BAO against different models.
The first row is the fit to the BAO using
a different parametrization of the peak, as a shift along the line-of-sight
and across: $(\alpha_{\parallel},\alpha_{\perp})$. In the second row
the BAO scale is fixed to its fiducial value: $\alpha_{\mathrm{iso}}=1$.
In the third row, we leave free the parameters setting the non-linear
broadening of the BAO peak, $(\Sigma_{\parallel},\Sigma_{\perp})$.
Finally, in the fourth row we fit for the size of the BAO peak
by leaving free the parameter $A_{\mathrm{peak}}$ from its fiducial value
and fix it to zero in the last row to get a model without a BAO scale.

The third section of the table presents changes to the model that affect the
overall shape of the cross-correlation without modifying the BAO scale.
The rows in this section represent various modifications.
1) we leave free the growth-rate of structure, $f$.
2) we leave free the shared redshift-space distortion parameter
of the three magnesium transitions, $\beta_{\mathrm{Mg}}$.
3) we leave free the
parameter giving the shared redshift evolution of the bias of the three Mg
species, $\gamma_{\mathrm{Mg}}$.
4) we allow the parameter giving the systematic redshift
error, $\Delta r_{\parallel}$, to be different for galaxies and for quasars.
5) we allow the parameter giving the statistic redshift
error and the effect of non-linear quasar velocities, $\sigma_{v}$,
to be different for galaxies and for quasars.
6) we replace the Lorentzian smoothing from measurement
error of the quasar redshift by a Gaussian smoothing.
7) we fix to zero the two parameters giving the effect of
systematic and statistic errors in the measurement of quasar redshift.
8,9) we either model the cross-correlation by
a single transition, or model the MgII doublet by an effective MgII singlet.

The last section of the table gives changes to the fitting range or to the
data used.
In the first two rows we fit either negative or positive values of separation
along the line-of-sight, $r_{\parallel}$.
In the next three rows, we change the fitting range in absolute separation, $r$.
In the sixth line we fit the correlation function in a narrower fitting range,
without the BAO feature.
The next two lines show the consequences on the fit, when the non-diagonal
elements of the covariance matrices are neglected, on the standard fit and
on a fit without the BAO peak.
Finally, in the last row we remove the galaxy-MgII(8), quasar-MgII(8) and galaxy-MgII(10)
cross-correlations where the
correlation matrix had to be replaced with a neighboring correlation
matrix to be positive definite.

\begin{table*}
    \centering
    \scalebox{0.98}{
    \begin{tabular}{lllllllll}

$\mathrm{Analysis}$ &
$\alpha_{\mathrm{iso}}$ &
$b_{\mathrm{MgII(2796)}}$ &
$b_{\mathrm{MgII(2804)}}$ &
$b_{\mathrm{MgI(2853)}}$  &  
$\chi^{2}_{\min{}}/DOF, probability$      \\ 

&
&
$\left[ 10^{-4} \right]$ &
$\left[ 10^{-4} \right]$ &
$\left[ 10^{-4} \right]$ &  
\\

\noalign{\smallskip}
\hline \hline
\noalign{\smallskip}

$\mathrm{Std.}       $ & $0.997 \pm 0.037$    & $\,\,\, -7.32 \pm 0.57$     & $\,\,\, -5.28 \pm 0.58$     & $-1.18 \pm 0.21$     & $75597.92 / (75120-6),  p = 0.11$ \\ 

\noalign{\smallskip}
\hline
\noalign{\smallskip}


$(\alpha_{\parallel},\alpha_{\perp})$ & $-$           & $\,\,\, -7.32 \pm 0.57$     & $\,\,\, -5.28 \pm 0.58$     & $-1.18 \pm 0.21$     & $75597.89 / (75120-7),  p = 0.11$ \\ 
$\,\,$ $(1.009,0.988) \pm (0.075,0.063)$ & & & & & \\

$\alpha_{\mathrm{iso}}=1$ & $1$        & $\,\,\, -7.32 \pm 0.57$     & $\,\,\, -5.28 \pm 0.58$     & $-1.18 \pm 0.21$     & $75597.93 / (75120-5),  p = 0.11$ \\ 

$(\Sigma_{\parallel},\Sigma_{\perp})$ & $0.997 \pm 0.037$    & $\,\,\, -7.29 \pm 0.55$     & $\,\,\, -5.31 \pm 0.55$     & $-1.19 \pm 0.21$     & $75597.90 / (75120-8),  p = 0.11$ \\ 
$\,\,$ $(0.3,0.0) \pm (1.3,1.9)$ & & & & & \\

$A_{\mathrm{peak}}$ & $1.000 \pm 0.022$    & $\,\,\, -7.29 \pm 0.54$     & $\,\,\, -5.28 \pm 0.53$     & $-1.18 \pm 0.21$     & $75592.41 / (75120-7),  p = 0.11$ \\ 
$\,\,$ $2.92 \pm 0.82$ & & & & & \\

$A_{\mathrm{peak}}=0 $ & $-$        & $\,\,\, -7.30 \pm 0.55$      & $\,\,\, -5.26 \pm 0.55$     & $-1.18 \pm 0.21$     & $75605.18 / (75120-5),  p = 0.10$ \\ 

\noalign{\smallskip}
\hline
\noalign{\smallskip}

$f$ & $0.997 \pm 0.037$    & $\,\,\,\,\,\, -7.3 \pm 3.9$       & $\,\,\,\,\,\,  -5.3 \pm 2.7$       & $-1.18 \pm 0.63$     & $75597.92 / (75120-7),  p = 0.11$ \\ 
$\,\,$ $0.79 \pm 0.41$ & & & & & \\

$\beta_{\mathrm{Mg}}$ & $0.997 \pm 0.036$    & $\,\,\, -8.75 \pm 0.78$     & $\,\,\, -6.04 \pm 0.61$     & $-1.44 \pm 0.25$     & $75587.92 / (75120-7),  p = 0.11$ \\
$\,\,$ $0.053 \pm 0.088$ & & & & & \\

$\gamma_{\mathrm{Mg}}$ & $1.002 \pm 0.037$    & $\,\,\, -6.89 \pm 0.52$     & $\,\,\, -5.04 \pm 0.48$     & $-1.20 \pm 0.21$      & $75570.81 / (75120-7),  p = 0.12$ \\ 
$\,\,$ $2.81 \pm 0.27$ & & & & & \\


$(\Delta r_{\parallel,\mathrm{galaxy}},\Delta r_{\parallel,\mathrm{quasar}})$ & $0.997 \pm 0.036$    & $\,\,\, -7.35 \pm 0.49$     & $\,\,\, -5.28 \pm 0.45$     & $-1.20 \pm 0.21$      & $75595.70 / (75120-7),  p = 0.11$ \\ 
$\,\,$ $(0.06,-0.64) \pm (0.34,0.51)$ & & & & & \\

$(\sigma_{v,\mathrm{galaxy}},\sigma_{v,\mathrm{quasar}})$ & $0.997 \pm 0.037$    & $\,\,\, -7.32 \pm 0.56$     & $\,\,\, -5.28 \pm 0.59$     & $-1.18 \pm 0.22$     & $75597.92 / (75120-7),  p = 0.11$ \\ 
$\,\,$ $(0.2,0.1) \pm (2.3,2.7)$ & & & & & \\

$\sigma_{v} \, \mathrm{Gauss}$ & $0.997 \pm 0.037$    & $\,\,\, -7.30 \pm 0.57$      & $\,\,\, -5.31 \pm 0.59$     & $-1.19 \pm 0.21$     & $75597.90 / (75120-6),  p = 0.11$ \\ 
$\,\,$ $0.5 \pm 2.0$ & & & & & \\

$(\Delta r_{\parallel},\sigma_{v}) = (0,0)$ & $0.997 \pm 0.037$    & $\,\,\, -7.26 \pm 0.36$     & $\,\,\, -5.34 \pm 0.26$     & $-1.18 \pm 0.21$     & $75597.96 / (75120-4),  p = 0.11$ \\ 
$(b_{\mathrm{MgII(2804)}},b_{\mathrm{MgI(2853)}})=(0,0)$ & $0.994 \pm 0.036$    & $-12.98 \pm 0.55$    & $-$        & $-$        & $75682.11 / (75120-4),  p = 0.072$ \\ 
$b_{\mathrm{MgII(2796)}}=0$ & $0.997 \pm 0.036$    & $-$        & $-13.08 \pm 0.57$    & $-1.40 \pm 0.29$      & $75649.21 / (75120-5),  p = 0.084$ \\ 

\noalign{\smallskip}
\hline
\noalign{\smallskip}

$r_{\parallel}<0$         & $1.225 \pm 0.076$    & $\,\,\, -10.0 \pm 1.8$      & $\,\,\, -1.2\,\,\,  \pm 2.2$ & $\,\,\,\,11.6 \pm 4.2$ & $38225.32 / (37560-6),  p = 0.0074$ \\ 
$r_{\parallel}>0$         & $0.953 \pm 0.056$    & $\,\,\, -6.84 \pm 0.65$     & $\,\,\, -5.34 \pm 0.65$      & $-1.19 \pm 0.22$       & $37385.31 / (37560-6),  p = 0.73$ \\ 
$r \in [0,160]$           & $0.997 \pm 0.037$    & $\,\,\, -7.02 \pm 0.25$     & $\,\,\, -5.54 \pm 0.24$      & $-1.22 \pm 0.21$       & $75875.51 / (75360-6),  p = 0.090$ \\ 
$r \in [40,160]$          & $0.996 \pm 0.038$    & $\,\,\, -7.4\,\,\, \pm 2.8$ & $\,\,\, -5.2\,\,\, \pm 2.8$  & $-1.40 \pm 0.36$       & $71165.15 / (70620-6),  p = 0.072$ \\ 
$r \in [10,180]$          & $0.996 \pm 0.037$    & $\,\,\, -7.38 \pm 0.56$     & $\,\,\, -5.29 \pm 0.59$      & $-1.19 \pm 0.21$       & $96113.97 / (95400-6),  p = 0.050$ \\ 
$r \in [40,160] + A_{\mathrm{peak}}=0$           & $-$                         & $\,\,\,-8.5 \,\,\,\pm 2.9$   & $\,\,\,-4.0 \,\,\,\pm 2.8$          & $-1.40 \pm 0.37$   & $71171.24 / (70620-5),  p = 0.070$ \\
$\mathrm{only\,diagonal}$                        & $0.960 \pm 0.026$           & $\,\,\,-5.76 \pm 0.42$       & $\,\,\,-4.3 \,\,\,\pm 0.40$         & $-1.17 \pm 0.13$   & $76111.28 / (75120-6),  p = 0.0052$ \\
$\mathrm{only\,diagonal}+ A_{\mathrm{peak}}=0$   & $-$                         & $\,\,\,-5.72 \pm 0.42$       & $\,\,\,-4.26 \pm 0.40$        & $-1.17 \pm 0.13$   & $76142.83 / (75120-5),  p = 0.0041$ \\
$\mathrm{only\,Pos.Def.}$                        & $1.004 \pm 0.038$           & $\,\,\, -7.39 \pm 0.53$      & $\,\,\, -5.23 \pm 0.47$ & $-1.22 \pm 0.22$   & $68070.75 / (67608-6),  p = 0.10$ \\

\noalign{\smallskip}

    \end{tabular}
    }
    \caption{
    Best fit results for the BAO parameter and the magnesium biases,
    for the combined fit to all 30 different cross-correlations
    under different assumptions in the analysis.
    The first section reproduces the last row of table~\ref{table::result_fit}
    for comparison.
    The second section gives changes in the BAO model.
    The third section gives changes in the cross-correlation model.
    The last section lists changes in the fitting range.
    When the analysis has extra free parameters, their best fits are given
    in the first column in parentheses immediately below the model description.
    }
    \label{table::result_fit_systematics}
\end{table*}

\end{document}